\documentclass[conference]{IEEEtran}
\IEEEoverridecommandlockouts
\usepackage{cite}
\usepackage{amsmath,amssymb,amsfonts}
\usepackage{graphicx}
\usepackage{textcomp}
\usepackage{xcolor}
\def\BibTeX{{\rm B\kern-.05em{\sc i\kern-.025em b}\kern-.08em
    T\kern-.1667em\lower.7ex\hbox{E}\kern-.125emX}}

\usepackage{algorithm}
\usepackage[noend]{algpseudocode}
\usepackage{subfigure}
\usepackage{footnote}
\usepackage{multirow}
\usepackage{booktabs}
\usepackage{enumitem}

\usepackage{amsmath}               
\usepackage{amsthm}

\newtheorem{assumption}{Assumption}

\newtheorem{definition}{Definition}

\newtheorem{theorem}{Theorem}

\newtheorem{lemma}{Lemma}

\numberwithin{table}{section}

\def\model{ASO-Fed}

\begin{document}

\title{Asynchronous Online Federated Learning for Edge Devices with Non-IID Data}


 
\author{
Yujing Chen,\textsuperscript{1} Yue Ning,\textsuperscript{2} Martin Slawski,\textsuperscript{3}  Huzefa Rangwala\textsuperscript{1}\\
\textsuperscript{1}{\textit{Department of Computer Science, George Mason University}}\\
\textsuperscript{2}{\textit{Department of Computer Science, Stevens Institute of Technology}}\\
\textsuperscript{3}{\textit{Department of Statistics, George Mason University}}\\
Email: ychen37@gmu.edu, yue.ning@stevens.edu, mslawsk3@gmu.edu, rangwala@gmu.edu
}

\maketitle

\begin{abstract}
Federated learning (FL) is a machine learning paradigm where a shared central model is learned across distributed devices while the training data remains on these devices. Federated Averaging (\textit{FedAvg}) is the leading optimization method for training non-convex models in this setting with a synchronized protocol. 
  However, the assumptions made by FedAvg are not realistic given the heterogeneity of devices. 
  First, the volume and distribution of collected data vary in the training process due to different sampling rates of edge devices.  
   Second, the edge devices themselves also vary in latency and system configurations, such as memory, processor speed, and power requirements. This leads to vastly different computation times. 
Third,
   availability issues at edge devices can lead to a lack of contribution from specific edge devices to the federated model.
  In this paper, we present an \textbf{As}ynchronous \textbf{O}nline \textbf{Fed}erated Learning (\model) framework, where the edge devices perform online learning with continuous streaming local data and a central server aggregates model parameters from clients. 
  Our framework updates the central model in an asynchronous manner to tackle the challenges associated with both varying computational loads at heterogeneous edge devices and edge devices that lag behind or dropout.
  We perform extensive experiments on a benchmark image dataset and three real-world datasets with non-IID streaming data. The results demonstrate \model~converging fast and maintaining good prediction performance.
\end{abstract}


\section{Introduction}
As massive data is generated from modern edge devices (e.g., mobile phones, wearable devices, and GPS),
distributed model training over a large number of computing nodes has become essential for many machine learning applications. 
With increasing popularity and computation power of these edge devices, federated learning (FL) has emerged as a potentially viable solution to enable the training of statistical models locally on the devices~\cite{mcmahan2016communication,konevcny2016federated2,konevcny2016federated}. 
FL involves training 
a shared central model from a federation of
distributed devices under the coordination of a central server; while the training data is kept on the edge device. Each edge device performs training on its local data and updates model parameters to the server for aggregation. 
Many applications can leverage this FL framework such as learning activities of mobile device users, forecasting weather pollutants, and predicting health events. 

\begin{figure}[t]
\centering
\includegraphics[trim={4.3cm 7.2cm 9.2cm 7.5cm}, clip,width=8cm]{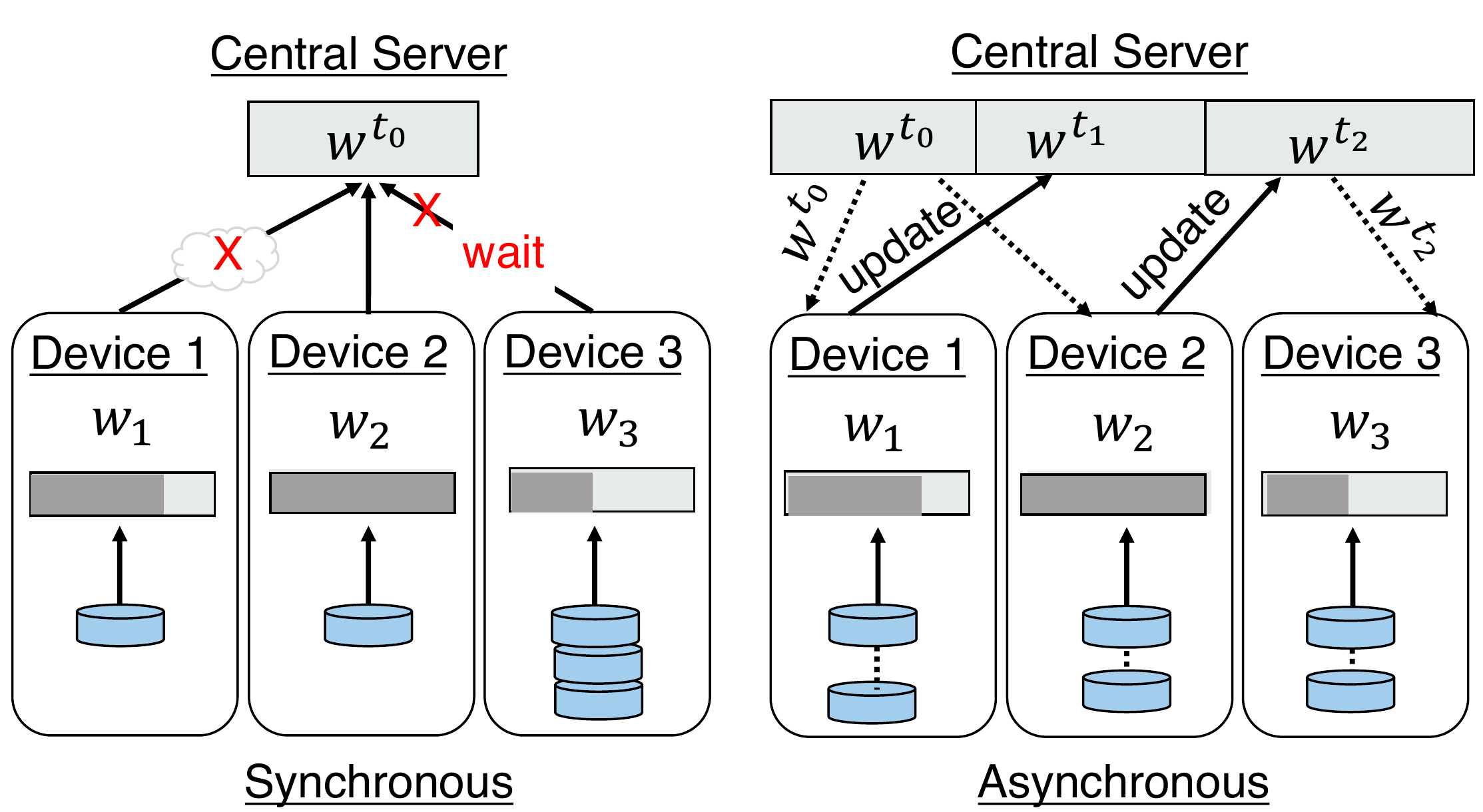}
	\caption{Illustration of Synchronous vs. Asynchronous update. In synchronous optimization, Device 1 has no network connection and Device 3 needs more computation time, thus the central server has to wait. Asynchronous updates do not need to wait.  
	}\label{fig:sync-async}
	\vspace{-2em}
\end{figure}

Many prior FL approaches use a synchronous protocol (e.g., \textit{FedAvg} \cite{mcmahan2016communication} and its extensions \cite{hard2018federated,konevcny2016federated,konevcny2016federated2,wang2018edge,leroy2019federated}), where at each global iteration, the server distributes the central model to a selected portion of clients and aggregates all updates from these clients by applying a weighted averaging strategy. 
These methods are costly due to a  synchronization \cite{chen2018lag} step (shown in Figure \ref{fig:sync-async}), where the server needs to wait for all local updates before aggregation. The existence of lagging devices (i.e., stragglers, stale workers) is inevitable due to device heterogeneity and network unreliability. To address this problem, asynchronous federated
learning methods \cite{xie2019asynchronous,chen2019efficient} were proposed, where the server can aggregate without waiting for the lagging devices. However, these asynchronous frameworks assume that the number of data samples on each device will not change during the training process, which is not practical in real-life settings. Data on local devices may increase during training, since sensors on these distributed devices usually have a high sampling frequency. In addition, Non-IID (i.e., not independent and identically distributed) and highly imbalanced characteristics of device data create challenges for effective model training \cite{smith2017federated}. In this paper, we focus on the question: \textit{Can we develop an asynchronous online federated learning framework with a convergence rate guarantee while maintaining an optimal prediction performance?}


We propose an asynchronous online federated learning framework (\model), where the central model does not wait for collecting and aggregating the gradient information from lagging clients, and clients perform online learning to deal with local streaming data. By design, \model~enables wait-free computation and communication, which ensures the model always converges better than synchronized FL frameworks. 
This study focuses on improving prediction performance and computation efficiency in FL instead of communication costs or privacy issues.
\model~shares similar privacy benefits as other general FL algorithms \cite{mcmahan2016communication,konevcny2016federated2,konevcny2016federated,hard2018federated} as the data does not leave the edge devices. 

In practice, we find that \model~is particularly useful for streaming data with heterogeneous devices having different  computing/communication speeds. Besides the prediction performance, we also simulate different network delays for each device to show the computational efficiency of \model. 
We summarize the main contributions of this paper as follows:

\begin{itemize}[leftmargin=*]
    \item We propose an asynchronous federated learning framework (\model) under a non-IID setting that allows updates from clients with continuously arriving data. The proposed framework learns inter-client relatedness effectively using regularization and a central feature learning module. We provide theoretical guarantees for the convergence of this proposed model. 
    \item We design a novel online learning procedure with a decay coefficient to balance the previous and current gradients on clients when handling streaming data. 
    \item 
    We introduce a dynamic learning strategy for step size adaptation on local devices to mitigate the impact of stragglers on the convergence of the central model. We show empirically that \model~is robust against data heterogeneity and network connections with high communication delays between the server and some clients.  
 
    \item We conduct extensive experiments on real-world and benchmark datasets, including a comparison with other state-of-the-art models. The results demonstrate that the proposed method achieves competitive prediction performances and converges fast with lower computation cost. 

\end{itemize}

\section{Related Work}

\subsection{Distributed Optimization}
As massive data are generated from edge devices such as mobile phones, wearable devices, and sensors, the computing power on these devices is also growing rapidly. Learning models directly on these distributed devices is gaining an increasing amount of attention.
Multi-task learning models are not suitable for edge device training given that they assume all clients (devices) participate in each training round. This requires that all clients are available because each client is training an individual specific model \cite{kairouz2019advances}. However, edge devices could be frequently offline during the training process due to unreliable networks or other factors. Parameter servers, where server nodes maintain globally shared parameters with data distributed on local nodes, however, often suffer from problems such as high network
bandwidth or communication overhead \cite{li2014communication,li2013parameter,mcmahan2016communication}.
Due to problems with stragglers and the non-IID character of edge device data, numerous other distributed optimization methods \cite{iutzeler2013asynchronous,aybat2015asynchronous,zhang2013communication,li2014communication,zhang2015deep,smith2017cocoa} in recent years are also not suitable for on-device learning. Federated Learning provides a promising solution that is  capable of dealing with heterogeneous devices across adhoc communication networks. 


Federated optimization methods have shown significant improvements on balancing communication versus computation  over traditional distributed approaches \cite{boyd2011distributed,dekel2012optimal}. Federated learning was first introduced by McMahan {\em et al.} \cite{mcmahan2016communication} and has been benchmarked on image and language datasets. 
Many extensions have been explored based on this original federated learning setting \cite{hard2018federated,nishio2019client,konevcny2016federated,konevcny2016federated2,caldas2018leaf}. 
A better approach to deal with non-IID data distribution is proposed by sharing a small amount of data with other devices \cite{zhao2018federated}. 
However, all these studies update the federated model in a synchronous fashion and do not tackle the problem of stragglers and dropouts.

Smith {\em et al.} \cite{smith2017federated} developed a primal-dual optimization method within a multi-task learning paradigm. This involved learning separate models for each device and dealing with stragglers. However,  this approach was not suitable for  non-convex formulations (e.g., deep learning), where strong duality is no longer guaranteed. 
Xie {\em et al.} \cite{xie2019asynchronous} proposed an asynchronous update procedure for federated optimization by updating the central model with weighted averaging, but this did not consider real-world scenarios where edge devices faced continuous streaming data.
%
Our proposed model assumes no constraints on the server aggregation procedure and obtains the optimal prediction performance on clients with heterogeneous data. In addition, we also incorporate online learning on clients to leverage the continuous arrival of new data points. 


\subsection{Online Learning with Multiple Clients}
Online learning methods operate on a group of data instances that arrive in a streaming fashion.
Most existing work in online learning across multiple clients are within the multi-task learning paradigm. Each client seeks to learn an individual model, in conjunction with related clients. 
The online learning problem with multiple tasks was first introduced by Dekel {\em et al.} \cite{dekel2006online}. The relatedness of participated tasks is captured by a global loss and aims to reduce the cumulative loss over multiple rounds. 
To better model task relationships, Lugosi {\em et al.} \cite{cavallanti2010linear}  impose a hard constraint on $K$ simultaneous
actions taken by the learner in the expert setting; Agarwal {\em et al.} \cite{agarwal2008matrix} use matrix regularization and Murugesan {\em et al.} \cite{murugesan2016adaptive} learn task relationship matrix automatically from the data. All these methods are proposed with synchronized protocols and not adaptable for real-world asynchronous learning.  

Jin {\em et al.} \cite{jin2015collaborating} developed a distributed framework to perform local training and central learning alternatively with a soft confidence-weighted classifier.
Although this is an asynchronous approach, it assumes that the local data is normally distributed, which is restrictive for non-convex neural network objectives. Besides, it lacks theoretical convergence guarantees and also requires each client to send a portion of its local data to the server.    

Different from the above online learning approaches, we design an iterative local computation procedure to balance the previous and current gradients. Besides, we also implement a local constraint to limit the deviation of local models from the central model and aim to learn an optimal central solution.



\section{Definitions and Preliminaries}
In this section, we first present the general form of federated learning. Then we briefly introduce the commonly used \textit{FedAvg}~\cite{mcmahan2016communication} and identify the issues in synchronized federated settings.


Assume that we have $K$ distributed devices. Let $\mathcal{D}_k$ denote data captured on device $k$, and define $n_k = |\mathcal{D}_k|$ to be the number of samples on device $k$. We denote $N = \sum_{k=1}^K |\mathcal{D}_k|$ as the total number of samples in $K$ devices. Assuming for any $k\neq k^{'}$, $\mathcal{D}_k\bigcap\mathcal{D}_{k^{'}} = \varnothing$. We then define local empirical loss of client $k$ as:
\begin{equation}
f_k(w_k) \overset{def}= \frac{1}{n_k}\sum_{i\in\mathcal{D}_k}^{} \ell_i(x_i,y_i;w_k).
\end{equation}
where $\ell_i(x_i,y_i;w_k)$ is the corresponding loss function for data point $\{x_i,y_i\}$ and $w_k$ is the local model parameter.
We can obtain the following central objective function: 
\begin{equation}\label{formula:syn-obj}
F(w) = \sum_{k=1}^{K}\frac{n_k}{N}f_k(w). 
\end{equation}
where $w$ is the aggregated central model.\footnote{We use $w$ and $w_k$ to represent the central model and client model, respectively} The overall goal is to find a model $w_*$ with: 
\begin{equation}
w_* = \arg \min F(w).
\end{equation}

\subsection{Synchronized Federated Optimization}

As shown in Algorithm \ref{alg:FedAvg}, for \textit{FedAvg}, at each global iteration, a subset of the devices are selected to run gradient descent optimization (e.g., SGD) locally to optimize the local objective function $f_k$ on device $k$. Then these local devices communicate their local model updates to the server for aggregation. 
With heterogeneous local objectives $f_k$, carefully tuning of local epochs is crucial for \textit{FedAvg} to converge. However, a larger number of local epochs may lead each device towards the optima of its local objective as opposed to the central objective. Besides, data continue to be generated on local devices which increases local gradient variations relative to the central model.  
Therefore, we incorporate a constraint to restrict the amount of local deviation by penalizing large changes from the current model at the server. We explain this in  detail in Section \ref{local-learning}. 

\begin{algorithm}[t]
\caption{Algorithm for FedAvg}\label{alg:FedAvg}
\begin{algorithmic}[1]
    \State \textbf{Input:} $K$ indexed by $k$, local minibatch size $B$, local epochs $E$ and learning rate $\eta$.
    \State \textbf{Central Server:}
    \For{global iterations $t=1,2,...,T$} 
        \State
        Server chooses a subset $S_t$ of $K$ devices at random 
        \For{each client $k \in S_t$ in parallel}
         \State $w_k^{t+1} \leftarrow$ ClientUpdate($k, w^t$ )
        \EndFor
     \State $w^{t+1} \leftarrow \sum_{k=1}^K \frac{n_k}{N} w_k^{t+1}$
   \EndFor
   \State \textbf{ClientUpdate($k,w^t$):}
   \State device $k$ updates $w^t$ for $E$ epochs of SGD on $f_k$ with $\eta$
   \State return $w_k^{t+1}$ to server
\end{algorithmic}
\end{algorithm}

Most synchronized federated optimization methods
have a similar update structure as \textit{FedAvg}. One apparent disadvantage of this structure is that, at each global iteration, when one or more clients are suffering from high network delays or clients which have more data and need longer training time, all the other clients must wait. Since the server aggregates after all clients finish, the extended period of waiting time in a synchronized optimization protocol will lead to idling and wastage of computing resources \cite{chai2019towards,chai2020tifl}.

\section{Proposed Method}

 We propose to perform asynchronous online federated learning where the server begins to update the central model $w$ after receiving updates from one client, without waiting for the other clients to finish. 
 The server maintains the current central model $w$, while all clients maintain their own copies ($w_k$) of $w$ in the memory. Note that the copy of $w$ at one client may be different from the copies at other clients.

\begin{figure}[t]
	\begin{center}
\includegraphics[trim={6cm 2cm 6cm 3.6cm}, clip,width=8.5cm]{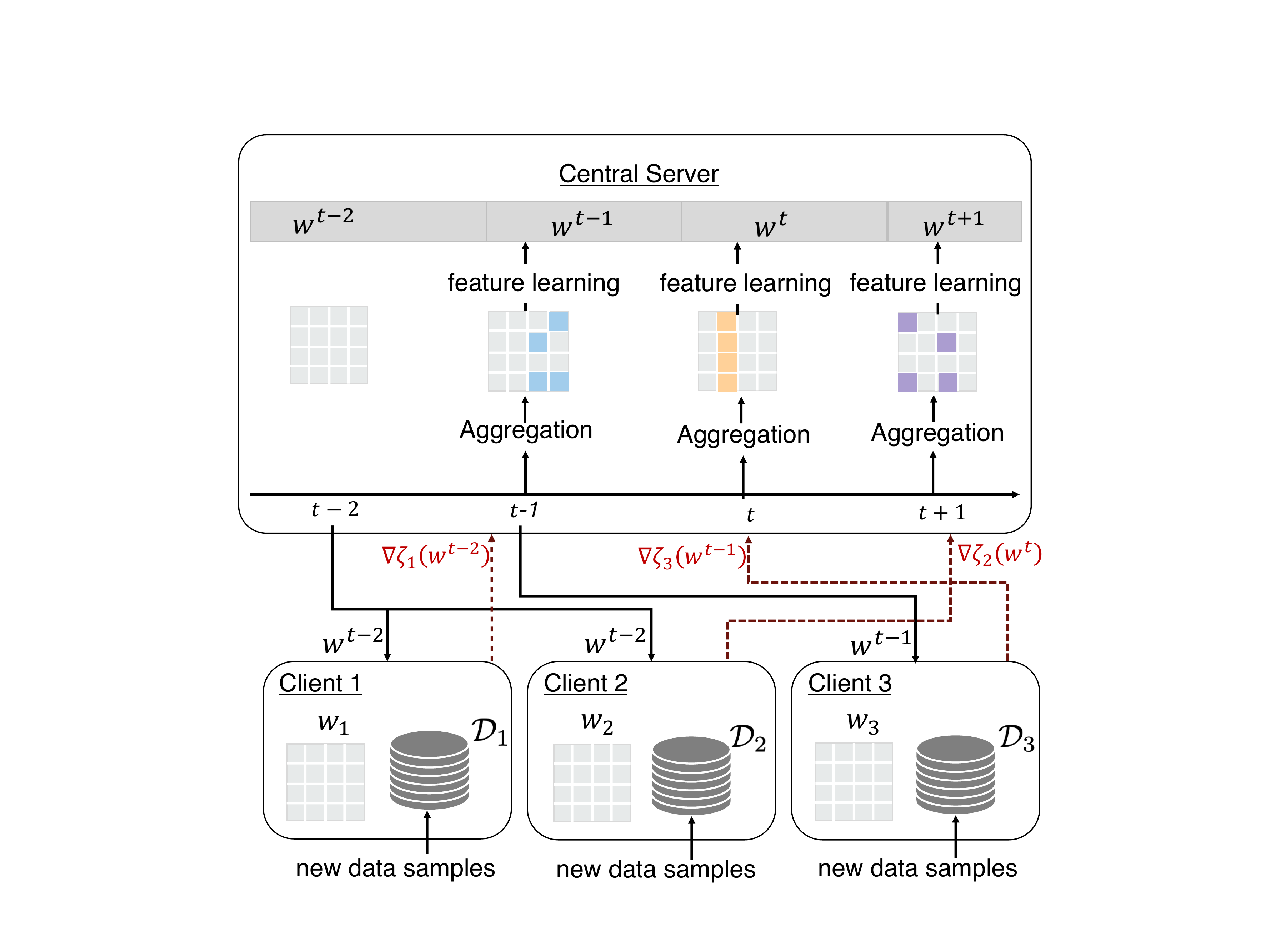}
	\caption{Illustration of update procedure for the proposed \textbf{\model} model. Server aggregates after receiving update from one client, and local clients may have new data samples during the training process. Each $w$ is used to represent the whole central/local model, $\nabla\zeta$ is the gradient of local client. 
	}\label{fig:AsyFL}
	\end{center}
	\vspace{-2em}
\end{figure}

Figure \ref{fig:AsyFL} illustrates the update procedure for \model. The server starts aggregation after receiving one client's update, and performs feature learning on the aggregated parameters to extract a cross-client feature representation. Then the server starts the next iteration and distributes the new central model to the ready clients. Clients may have new data samples during the training process. To better capture the inter-client relatedness, we use a decay coefficient to balance the previous and current local gradients with an iterative local computation procedure. The approach of \model~is detailed in Algorithm~\ref{alg:AsyOnFL}. We will explain each part in detail in the following sections.

As an example in Figure \ref{fig:AsyFL}, when the server node receives the gradient uploaded from the lagging clients (e.g., Client 2), it has already updated the central model twice. 
We can observe that there is an inconsistency in the asynchronous update scheme when it comes to obtaining model parameters from the server. Such inconsistency is common in the real-world settings and is caused by data and system heterogeneity, or network delay. We address this problem by learning a global feature representation on the server and using a dynamic learning step size for training local clients.

\subsection{Learning on Central Server}

The server aggregates the central model $w$ after each global iteration. At global iteration $t+1$, assume the server receives an update from client $k$. Let  $w^{t+1}$ be the server model, $w_k^t$ be the local model of client $k$ at iteration $t$, $\nabla\zeta_k$ be the gradient on the local data of client $k$, $\eta_k^t$ be the learning rate of client $k$ and $N' = n_1 + \dots + n'_{k}+\dots+ n_K$ be the current total number of data samples where $n_k$ and $N$ becomes $n'_k$ and $N'$ due to new data at client k. By aggregating the update from client $k$, the server update is computed as follows: 
\begin{equation}\label{equa:server}
\begin{aligned}
    w^{t+1} &= w^t - \frac{n'_k}{N'}(w_k^t - w_k^{t+1})\\
    &= w^t -\frac{n'_k}{N'}( w_k^t - (w_k^t - \eta_k^t \nabla \zeta_k(w^t))) \\
    &= w^t - \eta_k^t\frac{n'_k}{N'} \nabla\zeta_k(w^t). 
\end{aligned}
\end{equation}
\noindent\textbf{Feature Representation Learning on Server.}
To address the potential effect on model performance caused by asynchronous updates, we apply feature representation learning on the server to extract a cross device feature representation. Attention mechanisms have shown to be effective in identifying key  features and their representations \cite{firat2016multi,vaswani2017attention}. Our feature learning approach is inspired by this, 
and additionally, we combine weight normalization to reduce the computation cost \cite{wang2006normalization,ba2016layer}. We use simple network architectures in this study so that it can be easily handled by mobile devices. We apply feature extraction on the first layer (e.g., LSTM or CNN in this paper) after the input to generate the feature representation, and denote the parameters of this layer as $w^{{t+1}}_{(1)}$. 
For each element $w^{{t+1}}_{(1)}[i,j]$ in column $w^{{t+1}}_{(1)}[j]$ of $w^{{t+1}}_{(1)}$, we adopt the below operations to obtain the updated $w^{t+1}_{(1)}$: 
\begin{align}\label{equa:feature1}
    \alpha^{{t+1}}_{(1)}[i,j] &\leftarrow \frac{\exp(|w^{{t+1}}_{(1)}[i,j]|)}{\sum_{j}^{}\exp(|w^{{t+1}}_{(1)}[i,j]|)} ,\\
\label{equa:feature2}
    w^{{t+1}}_{(1)}[i,j] &= \alpha^{{t+1}}_{(1)}[i,j] \ast w^{{t+1}}_{(1)}[i,j].
\end{align}

\subsection{Learning on Local Clients}\label{local-learning}

\begin{algorithm}[t]
\caption{Algorithm for \model}\label{alg:AsyOnFL}
\begin{algorithmic}[1]
    \State \textbf{Input:} Multiple related learning clients distributed at client devices, 
  regularization parameter $\lambda$, multiplier $r_k$, learning rate $\bar\eta$, decay coefficient $\beta$.
 \State \textbf{Initialize:} $h_k^{pre} = h_k = 0$, $v_{k}=0$
\State \underline{\textbf{Procedure at Central Server}} 
  \For{global iterations $t=1,2,...,T$} 
       \State  /* get the update on $w^t$ */
        \State
        compute $w^t$ \Comment{[Eq.(\ref{equa:server})]} 
       \State  update $w^t$ with feature learning \Comment{[Eq.(\ref{equa:feature1}) - Eq.(\ref{equa:feature2})]} 
   \EndFor
   \State  \textbf{end for}
\State \underline{\textbf{Procedure of Local Client $k$ at round $t$}}
     \State receive $w^t$ from the server
    
     \State Compute $\nabla s_k$
     \State Set $h_{k}^{\text{(pre)}} = h_k$
     \State Set $\nabla\zeta_k\leftarrow \nabla s_k - \nabla s_{k}^{\text{(pre)}} + h_{k}^{\text{(pre)}}$ 
      \Comment{[Eq.(\ref{local:object}) -Eq.(\ref{formula:local-up})]}

 \State Update $w_k^{t+1} \leftarrow w_k^t-r_k^t\bar\eta\nabla\zeta_k $
 \State Compute and update $h_k = \beta h_k + (1-\beta) v_k$
 \State Update $v_{k} = \nabla s_k(w^t;w_k^t)$
 \State upload $w_k^{t+1}$ to the server
\end{algorithmic}
\end{algorithm}

In order to mitigate the deviations of the local models from the central model, instead of just minimizing the local function $f_k$, device $k$ applies a gradient-based update using the following surrogate objective $s_k$:
\begin{equation}\label{local:object}
    s_k(w_k) =  f_k(w_k) + \frac{\lambda}{2} ||w_k-w||^2.
\end{equation}
\noindent\textbf{Local Update with Decay Coefficient.} Data continue arriving at local clients during the training process, so each client needs to perform online learning. For this process, each client requests the latest model from the server and updates the model with its new data. Thus there needs to be a balance between previous model and current model. At global iteration $t$, device $k$ receives model $w^t$ from the server. Let $\nabla s_{k}^{\text{(pre)}}$ be the previous local gradients, the optimization of device $k$ at this iteration is formulated as:
\begin{align}
    \nabla\zeta_k \leftarrow \nabla s_k - \nabla s_{k}^{\text{(pre)}} + h_{k}^{\text{(pre)}}, \\
    h_{k}^{\text{(pre)}} = \beta h_{k}^{\text{(pre)}}  + (1-\beta) \nabla s_{k}^{\text{(pre)}}.
\end{align}

where $h_{k}^{\text{(pre)}} $ is used to balance the previous and current local gradients and initialized to be $0$, $\beta$ is the decay coefficient to balance the previous model and the current model. The update procedure of $h_{k}^{\text{(pre)}} $ can be found in Algorithm \ref{alg:AsyOnFL}.


With $\eta_k^t$ being the learning rate for client $k$, the closed form solution for model update of client $k$ is given by:
\begin{equation}\label{formula:local-up}
\begin{aligned}
w_k^{t+1} &= w_k^t-\eta_k^t \nabla\zeta_k(w^t)  \\ 
      &= w_k^t- \eta_k^t\Big(\nabla f_k(w_k^t) - \nabla s_{k}^{\text{(pre)}} + h_{k}^{\text{(pre)}}  +\lambda (w_k^t-w^t)\Big).
\end{aligned}
\end{equation}

\noindent\textbf{Dynamic Learning Step Size.}
In real-world settings, the activation rates, i.e., how often clients provide updates to the central model, vary due to a host of reasons.  Devices with low activation rates are referred as stragglers, which are caused by several reasons such as communication bandwidth, network delay or data heterogeneity. Thus, we apply a dynamic learning step size with the intuition that if a client has more data or poor communication bandwidth, the activation rate of this client towards the central update will be small and thus the corresponding learning step size should be large. Dynamic learning step sizes are used in asynchronous optimization to achieve better learning performance \cite{cheung2014amortized,baytas2016asynchronous}. Initially, we set $\eta_k^t = \bar\eta$ for all clients. The update process (\ref{formula:local-up}) can be revised as:
\begin{equation}
    w_k^{t+1} = w_k^t - r_k^t\bar\eta \nabla\zeta_k(w^t).
\end{equation}

where $r_k^t$ is a time related multiplier, and is given by $r_k^t = \max\{1, \log(\bar d_k^t)\}$, where $\bar d_k^t = \frac{1}{t}\sum_{\tau=1}^{t}d_k^{\tau}$ is the average time cost of the past $t$ iterations. 
Then the actual learning step size is scaled by the past communication delays. This dynamic learning step size strategy can reduce the effect of stragglers on model convergence. 
Since the stragglers usually have longer delays, the larger step sizes are assigned to these lagging clients to compensate for the loss.

\subsection{Convergence Analysis}
In this section, we prove theoretical analysis on the convergence of \model. First, we introduce some definitions and assumptions for our convergence analysis. 

\begin{definition}
(Smoothness) The function $f$ has Lipschitz continuous gradients with constant $L > 0$ (in other words, $f$ is \textit{L-smooth}) if $\forall x_1,x_2$,
\begin{equation}
f(x_1)-f(x_2) \leq \langle\nabla f(x_2), x_1 - x_2\rangle + \frac{L}{2}||x_1 - x_2||^2.
\end{equation}
\end{definition}

\begin{definition}
(Strong convexity) The function $f$ is $\mu$-\textit{strongly convex} with $\mu > 0$ if $\forall x_1,x_2$,
\begin{equation}
f(x_1)-f(x_2) \geq \langle\nabla f(x_2), x_1 - x_2\rangle + \frac{\mu}{2}||x_1 - x_2||^2.
\end{equation}
\end{definition}

In order to quantify the dissimilarity between devices in a federated network, following Li {\em et al}  \cite{sahu2018federated}, we define the following definition on local non-IID data.

\begin{definition}\label{b-gradient}
(Bounded gradient dissimilarity): The local functions $\zeta_k$ are $V$-locally dissimilar at $w$ if $\mathbb{E}||(\nabla\zeta_k(w)) ||^2\leq ||\nabla F(w)||^2V^2$. 
\end{definition}

 With Definition \ref{b-gradient} we further define $V(w) = \sqrt{\frac{\mathbb{E}||\nabla F(w)||^2}{||(\nabla\zeta_k(w)) ||^2}}$ when $||(\nabla\zeta_k(w)) ||^2 \neq 0$. As explained in \cite{sahu2018federated}, when all the local functions are the same, which is the samples on all the devices are in IID fashion, we have $V(w) \rightarrow 1$ for all $w$. However, in federated setting with heterogeneous data, we often have $V > 1$ due to device discrepancies. Therefore, $V\geq 1$, and the larger $V$ is, the larger the dissimilarity among the local functions, which is the more heterogeneous the local data. 

Further, we make the following assumptions on the objective functions and introduce one lemma.

\begin{assumption}
\label{assumption1}
Suppose that:\\
1. The central objective function $F(w)$ is bounded from below, i.e., $F_{\min} = F(w_*) > -\infty$.\\ 
2. There exists $\epsilon > 0$ such that $\mathbb{E}(\nabla\zeta_k(w))\leq||\nabla F(w)||$, and $\nabla F(w)^\top \mathbb{E}(\nabla\zeta_k(w)) \geq \epsilon ||\nabla F(w)||^2$ holds for all $w$. 

\end{assumption}

Note that if $\epsilon = 1$, then $\nabla\zeta_k(w)$ is an unbiased estimator of $\nabla F(w)$.




\begin{lemma}
\label{lemma_1}
If $F(w)$ is $\mu$-\textit{strongly convex}, then with Assumption \ref{assumption1}.1, we have: 
\begin{equation}
2\mu(F(w^t) - F(w_*)) \leq  ||\nabla F(w^t)||^2.
\end{equation}
\end{lemma}

While the proof of \textbf{Lemma 1} is supported by the literature \cite{nesterov2013introductory,bottou2018optimization}, we also provide a detailed proof in Appendix A.


\begin{theorem}[]
\label{convex-model}
Suppose that the central objective function $F(w)$ is \textit{L-smooth} and $\mu$-\textit{strongly convex}. Assume the local functions $\zeta_k$ are bounded dissimilar. Let \textit{Assumption} \ref{assumption1} hold.  
Let $\bar\eta_k \leq \eta_k^t < \eta_k =  \frac{2\epsilon N'}{L V^2 n'_k}$, then after $T$ global
updates on the server, \model~converges to a global optimum $w_*$: 
\begin{equation}\label{Eq_4}
\begin{aligned}
\mathbb{E}(F(w^{T}) - F(w_*)) &\leq (1-2\mu\gamma'\bar\eta_k)^{T}(F(w^0) - F(w_*))
\end{aligned}
\end{equation}
where $\gamma' = \epsilon-\frac{L \eta_k V^2}{2}$.
\end{theorem} 

The detailed proof of Theorem \ref{convex-model} is provided in Appendix B. Theorem \ref{convex-model} converges under the special case of convex central loss and gives an error bound for the general form of model aggregation.


\begin{theorem}[]
\label{nonconvex-model}
Suppose that the central objective function $F(w)$ is  \textit{L-smooth} and non-convex.
Let \textit{Assumption} \ref{assumption1} hold. Assume the local functions $\zeta_k$ are bounded dissimilar. If it holds that $\eta_k^t < \frac{2\epsilon-1}{L V^2} \leq \max({r_k^t}\bar\eta) = \bar\eta$ for all $t$, then after $T$ global iterations, we have
\begin{equation}\label{Eq_nonconvex}
\begin{aligned}
\sum_{t=0}^{T-1} \frac{\eta_k^t}{2}\mathbb{E}(||\nabla F(w^t)||^2)
&\leq F(w^0)- F(w_*)
\end{aligned}
\end{equation}
\end{theorem}

We direct the reader to Appendix C for a detailed proof of Theorem~\ref{nonconvex-model}. The model convergence rate 
can be controlled with a balance between the bounded gradient dissimilarity value $V$ and the learning rate $\eta_k^t$.

\begin{table*}
\centering
\caption{Prediction performance comparison. Bold numbers are the best performance, numbers with underlines are the second best values. improv.(1) shows the percentage improvement of \model~over FedAvg. improv.(2) shows the percentage improvement of \model~over the best baseline results.} \label{table:performance}
\resizebox{\textwidth}{!}{\begin{tabular}{lccccccccccc}
\toprule
        \multicolumn{1}{c}{\multirow{3}{*}{Method}} & \multicolumn{4}{c}{\textbf{FitRec}}                                                         & \multicolumn{2}{c}{\textbf{Air Quality}}              & \multicolumn{4}{c}{\textbf{ExtraSensory}} & \textbf{Fashion-MNIST}                                                                 \\
       \cmidrule(r){2-5}\cmidrule(r){6-7}\cmidrule(r){8-11}\cmidrule(r){12-12}
\multicolumn{1}{c}{}                         & MAE $\downarrow$                 & SMAPE$\downarrow$             & MAE$\downarrow$                 & SMAPE$\downarrow$              & \multirow{2}{*}{MAE$\downarrow$} & \multirow{2}{*}{SMAPE$\downarrow$} & \multirow{2}{*}{F1$\uparrow$} & \multirow{2}{*}{Precision$\uparrow$} & \multirow{2}{*}{Recall$\uparrow$} & \multirow{2}{*}{BA$\uparrow$} & \multirow{2}{*}{Accuracy$\uparrow$}\\
\multicolumn{1}{c}{}                         & (Speed)             & (Speed)            & (HeartRate)         & (HeartRate)          &                      &                        &                     &                            &                         &   &                     \\ 
\midrule

FedAvg   & 13.61  & 0.78     & 13.72 & 0.78    & 44.30             & 0.44     & 0.66    & 0.87     & 0.55     & 0.77  & 0.87 \\
FedProx   & 14.21  &  0.82   & 14.53 &  0.83 &   44.30     & 0.44     & 0.67  & 0.82 & 0.57  &0.77  &  0.88\\

FedAsync   & 13.56  & 0.78   & 13.67 & 0.78  &        37.98     & $\underline{0.43}$   &  0.72 &  0.84  & 0.65 & 0.82  & 0.90 \\
Local-S &  12.76 &  0.75    & 13.27 &  0.76   &   $\underline{36.72}$   &  0.56  &  0.65  & 0.72  & 0.61 & 0.79 & 0.89\\

Global & 12.95  &  0.78   &12.79  & 0.79  & 37.61   & 0.44  & $\mathbf{0.77}$  & $\mathbf{0.92}$  & 0.66  & 0.83  & 0.92 \\ 
\midrule

\model(-D) &$\underline{12.46}$  & $\underline{0.74}$ &$\underline{12.51}$  & $\underline{0.75}$ & 37.13 & $\underline{0.43}$ & $\underline{0.76}$ & $\underline{0.88}$ & $\underline{0.69}$ & $\mathbf{0.85}$ & $\underline{0.94}$\\

\model(-F) & 12.62 &  0.76& 12.71 & 0.76  & 37.72 & $\underline{0.43}$  & 0.75 & 0.86 & 0.68  &$\underline{0.84}$  & $\underline{0.94}$\\

\model    & $\mathbf{12.31}$  &  $\mathbf{0.73}$  & $\mathbf{12.36}$ & $\mathbf{0.74}$  &  $\mathbf{36.71}$  & $\mathbf{0.42}$  & $\mathbf{0.77}$ & $\underline{0.88}$  &  $\mathbf{0.70}$  &  $\mathbf{0.85}$ & $\mathbf{0.95}$ \\ 

\midrule
improv.(1)    &9.55\%  & 6.41\%  & 9.91\%  &  5.13\%  & 17.13\% &  2.32\%  & 16.66\%  & 1.15\%  &  27.27\%  &  10.39\% & 9.19\%\\
improv.(2)    & 3.52\%  & 2.67\%  & 3.36\% &   2.63\% & 0.03\% &  4.54\%  &   0.00\%  &  -4.34\%  &  6.06\%   &  2.41\% & 3.26\% \\
\bottomrule
\end{tabular}}

\end{table*}

\begin{figure*}[h]
 \centering
\includegraphics[trim={1.5cm 0.8cm 0.9cm 0.8cm},clip, width = 16cm]{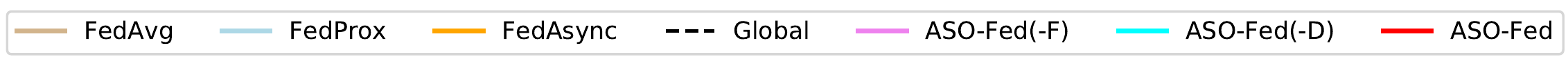}
\vspace{-1em}
\end{figure*}
\begin{figure*}[htbp]
  \centering
  \subfigure[FitRec (SMAPE $\downarrow$)]{\includegraphics[scale=0.27]{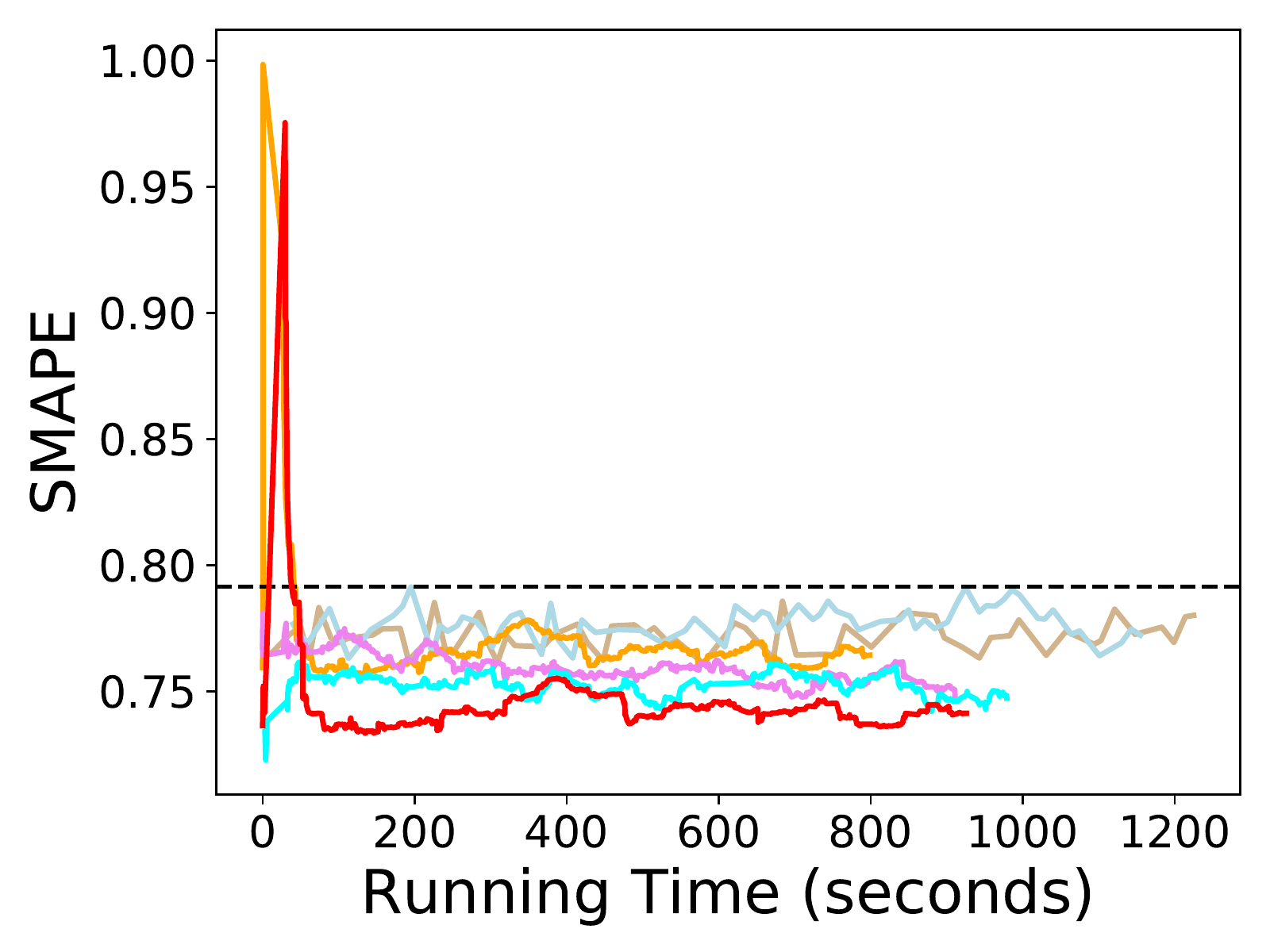}}
  \subfigure[Air Quality (SMAPE $\downarrow$)]{\includegraphics[scale=0.27]{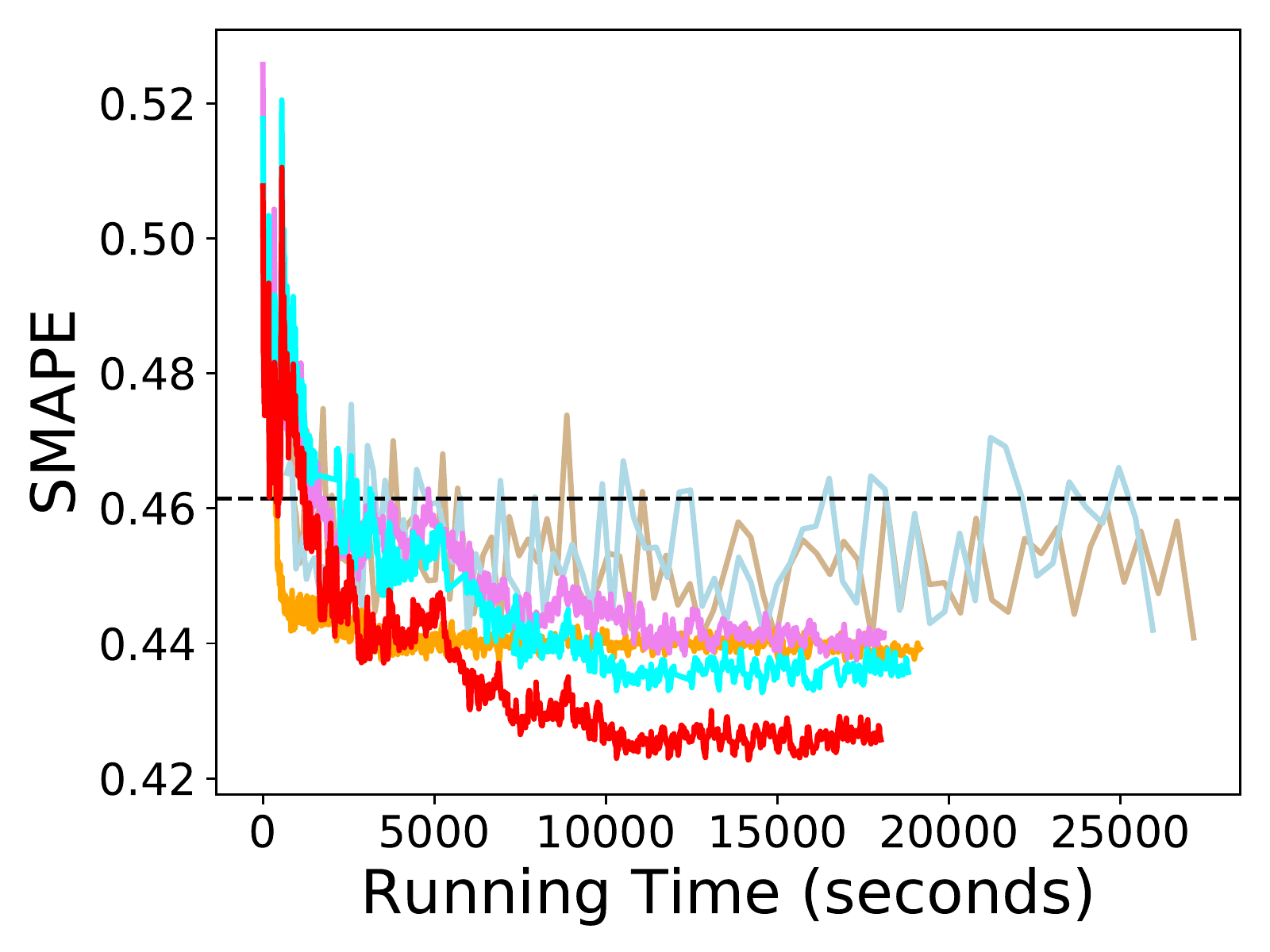}}
  \subfigure[ExtraSensory (F1-score $\uparrow$)]{\includegraphics[scale=0.27]{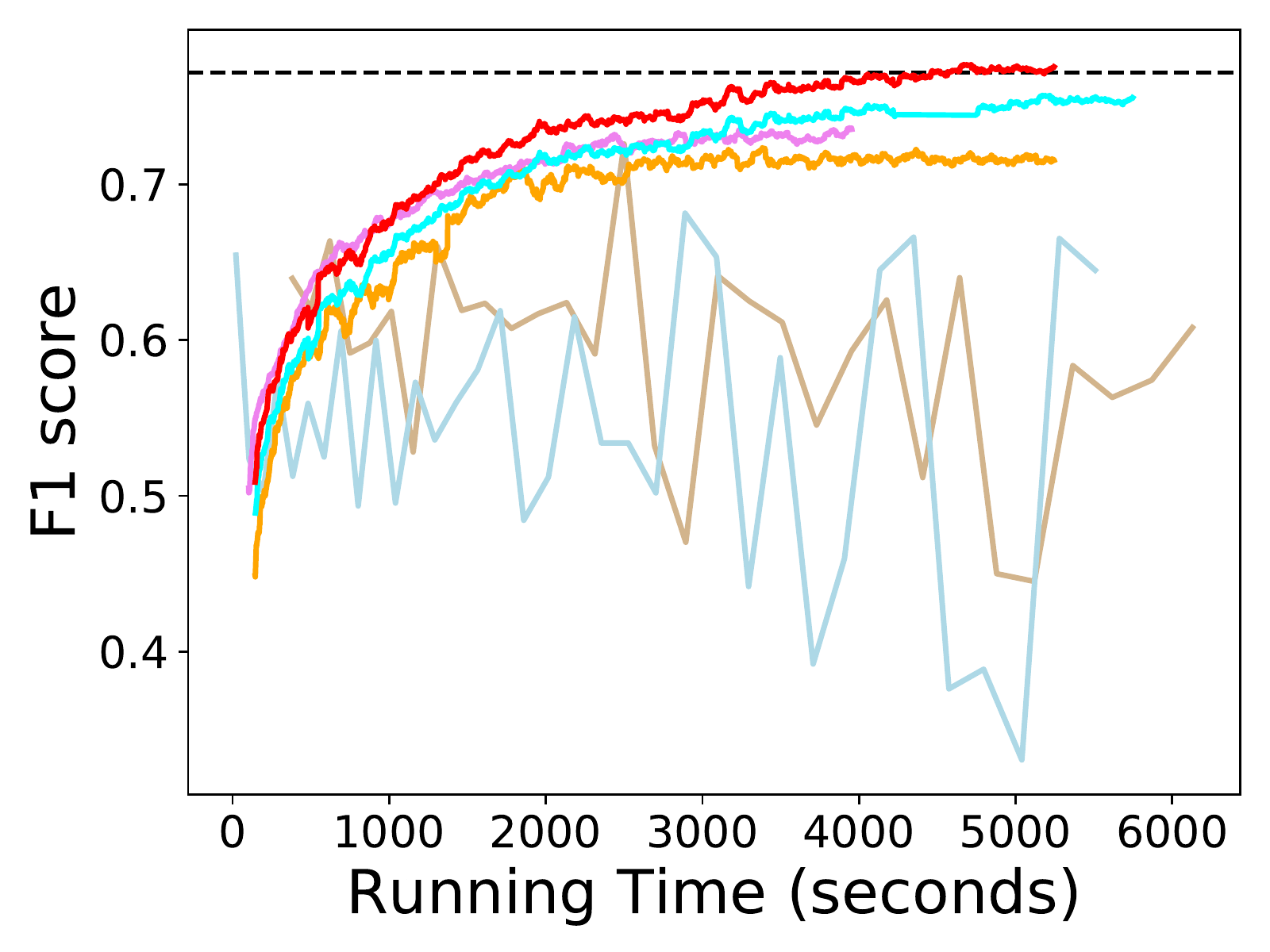}}
  \subfigure[Fashion-MNIST (Accuracy $\uparrow$)]{\includegraphics[scale=0.27]{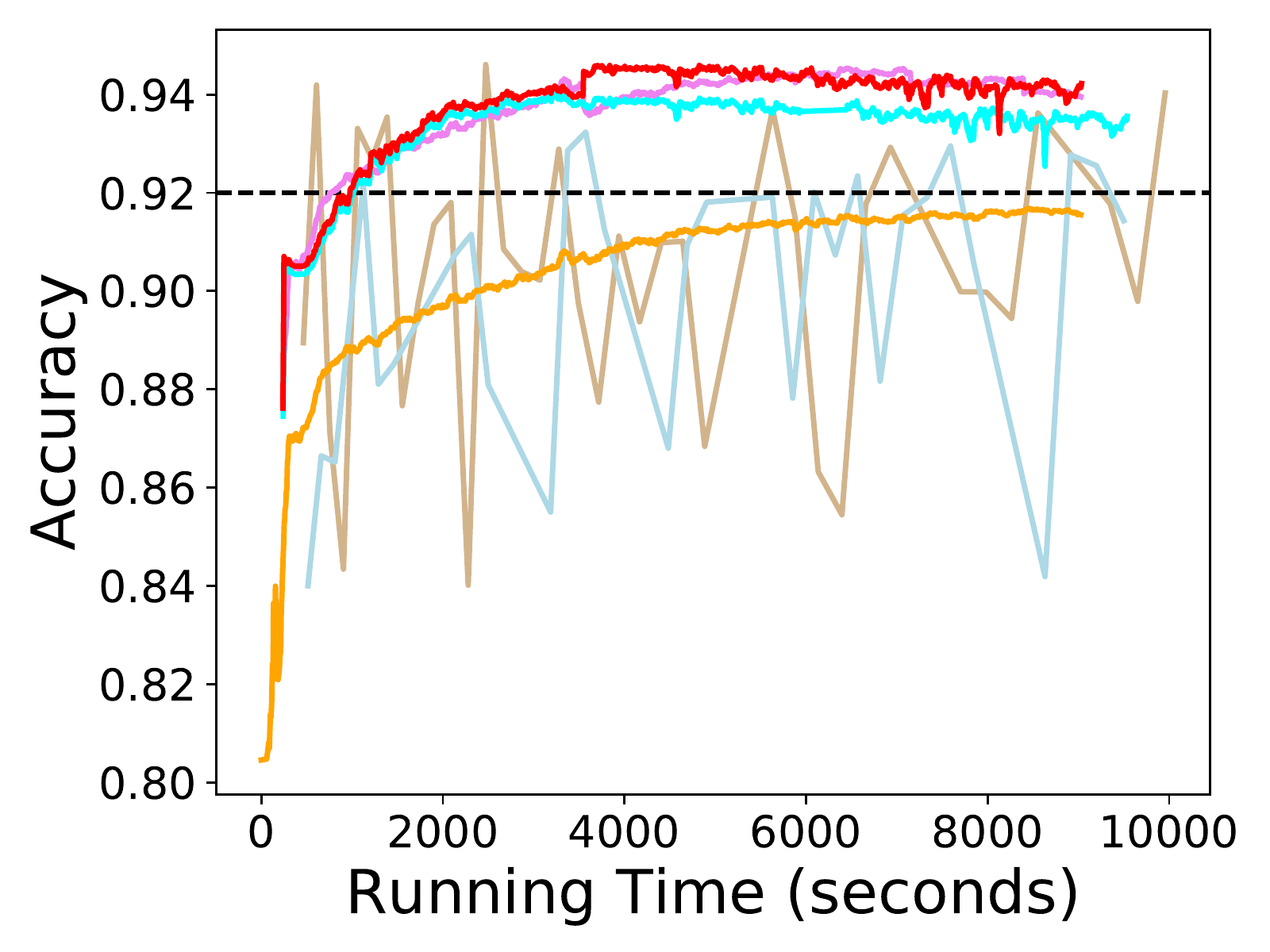}}
  \caption{Test set performance vs. running time for four datasets. Lower SMAPE value indicates better model performance. For the synchronized federated frameworks, we plot results of \textit{FedAvg} and \textit{FedProx} at every $10$ global iterations. }\label{tab:prediction-curve}
\end{figure*}

\section{Experimental Setup}

We perform extensive experiments on three real-world datasets and one benchmark dataset (Fashion-MNIST). 

\subsection{Datasets} 

\begin{itemize}[leftmargin=*]
    \itemsep0em 
    \item \textbf{FitRec Dataset\footnote{https://sites.google.com/eng.ucsd.edu/fitrec-project/home}:} 
    User sport records generated on mobile devices and uploaded to Endomondo, including multiple sources of sequential sensor data such as heart rate, speed, and GPS as well as the sport type (e.g., biking, hiking). Following \cite{ni2019modeling}, we re-sampled the data in 10-second intervals, and further generated two derived sequences: derived distance and derived speed. We use data of randomly selected 30 users for heart rate and speed prediction, and data of each user has features of one sport type. 
	\item \textbf{Air Quality Dataset\footnote{https://biendata.com/competition/kdd\_2018/data/}:} Air quality data collected from multiple weather sensor devices distributed in 9 locations of Beijing with features such as thermometer and barometer. Each area is modeled as a separate client and the observed weather data is used to predict the measure of
	six air pollutants (e.g., PM2.5). 
	\item \textbf{ExtraSensory Dataset\footnote{http://extrasensory.ucsd.edu/}:}  
	Mobile device sensor data (e.g., high-frequency motion-reactive sensors, location services, audio, watch compass) and watch sensor data (accelerator) collected from 60 users; performing any of 51 activities \cite{vaizman2018extrasensory}. 
	We use the provided 225-length feature vectors of time and frequency domain variables generated for each instance. We model device of each user as a client and predict their activities (e.g., walking, talking, running).
	\item \textbf{Fashion-MNIST:} This is a dataset of Zalando's article images—consisting of a training set of 60,000 examples and a test set of 10,000 examples. Each example is a 28x28 grayscale image, associated with a label from 10 classes (e.g., Dresses, Coat, Bag). Each class has the same number of examples. We follow a non-IID setting as in \cite{mcmahan2016communication} and divide the data into 20 parts according to their labels. We first sort the data by category label, divide each category into 4 different sizes $\{2000,2750,3250,4000\}$, and assign each of 20 parts 2 different sizes. We model each part as a separate client and predict the target labels.
\end{itemize}

\subsection{Comparative Methods} 
\subsubsection{Baseline Methods} We compare the proposed \model~with the following synchronous and asynchronous federated learning approaches, single-client and global models. 
\begin{itemize}[leftmargin=*]
    \itemsep0em 
    \item FedAvg \cite{mcmahan2016communication,konevcny2016federated2,konevcny2016federated}: the commonly used synchronous federated learning approach proposed by McMahan {\em et al.} \cite{mcmahan2016communication}.
    \item FedProx \cite{sahu2018federated}: synchronous federated learning framework with a proximal term on the local objective function to mitigate the data heterogeneity problem and to improve the model stability compared to FedAvg.
    \item FedAsync \cite{xie2019asynchronous}: asynchronous federated learning framework using a weighted average to update the server model.
    \item Local-S: each client learns separate model with its own data, and the model structure is the same as \model.
    \item Global: combining the data of all clients and processed in a batch setting on a single machine.
\end{itemize}
\subsubsection{Ablation Studies} We also perform ablation studies to study the effect of feature representation learning on server and local dynamic learning step size:
\begin{itemize}[leftmargin=*]
    \item \model(-D): the proposed \model~without dynamic learning step size.
    \item \model(-F): the proposed \model~without central feature representation learning.
\end{itemize}

\subsection{Training Details}

For each dataset, we split the each client's data into $60\%$, $20\%$, $20\%$ for training, validation, and testing, respectively. As for each client's training data, we start with a random portion of the total training size, and increase by $0.05\%-0.1\%$ each iteration to simulate the arriving data. We set the fraction $C$ of FedAvg as $0.2$, decay  coefficient $\beta$ as $0.001$, $\bar\eta = 0.001$, $\lambda = 1.0$ for FitRec and Air Quality datasets, $\lambda = 0.8$ for ExtraSensory dataset, and $\lambda = 0.5$ for Fashion-MNIST. For FedAsync model, we set $\gamma = 0.1$, $\rho = 0.005$ and $\alpha = 0.6$. We use a single layer LSTM followed by one fully connected layer for the three real-world datasets and two CNN layers followed by one Max Pooling layer for Fashion-MNIST. The local epoch number of each client is set as $2$. We design simple network architectures for all datesets so that it can be easily handled by mobile devices. 

\noindent\textbf{Simulation parameters.} To simulate the stragglers and dropouts situations, we set different network settings for our experiments, a random offset parameter ($10\sim100$ seconds) was taken as an input from the client. This parameter represents the average delay related to the infrastructure of the network for a client. 
We direct the reader to Section \ref{experiments} for the detailed results.

\section{Experimental Results}\label{experiments}

\subsection{Predictive Performance Comparison}

Table~\ref{table:performance} reports the predictive performance comparing \model~to the baseline approaches.  In case of  regression problems, we report the average \textit{MAE} and \textit{SMAPE} values, for ExtraSensory classification benchmark we report the average \textit{F1}, \textit{Precision}, \textit{Recall} and \textit{Balanced Accuracy (BA)}, and for Fashion-MNIST we report the \textit{Accuracy}.
%
From Table \ref{table:performance}, we observe that \model~achieves the lowest MAE and SMAPE values for FitRec and Air Quality datasets, and has the overall best performance for ExtraSensory and Fashion-MNIST datasets.
Across the four datasets, \model~outperforms 
the Global model (acquires all data at a single server)  by $2.39\% \sim 6.41\%$. 
FedAvg and FedProx do not perform  well on the highly unbalanced and non-IID datasets (FitRec, ExtraSensory and Fashion-MNIST). In particular, 
for the FitRec and Fashion-MNIST datasets, Local-S (the local single client model) outperforms FedAvg and FedProx. 

Figure \ref{tab:prediction-curve} shows the prediction performance for the different approaches as a function of running time.
From Figure \ref{tab:prediction-curve}, we notice large fluctuations on the performance of FedAvg and FedProx during the whole training process on all four benchmarks. This  shows that synchronous federated frameworks do not perform well on streaming data with skewed and non-IID data distribution.
FedAsync achieves better performance than the two synchronous federated frameworks, but not as good as \model. \model~has steady improvements with running time (and converges quickly). 

\begin{table}[h]
\small
\centering
\caption{Computation time (in minutes)
to reach target test performance. The network delay of each client was set to be a random value between $10\sim100$ seconds. }\label{tab:timecost}
\scalebox{0.9}{
\begin{tabular}{lcccc}
\toprule
Method & \textbf{FitRec} & \textbf{Air Quality} & \textbf{ExtraSensory}& \textbf{FMNIST} \\ 
\midrule
FedAvg         & 20.42 & 460.02 & 104.86  & 160.72 \\
FedProx         & 19.26 & 439.95 & 99.45
& 160.36 \\
FedAsync        & 15.41 &326.45  & 87.97  & 151.72 \\
\midrule
\model(-D)         &16.31 & 332.74 &  95.77 & 158.83 \\ 
\model(-F)         & $\mathbf{15.17}$ & 320.92 & $\mathbf{65.87}$ & 150.54\\
\model  & 15.43 & $\mathbf{319.41}$ &  87.40 & $\mathbf{150.46}$  \\ 
\bottomrule
\end{tabular}
}
\end{table}

\begin{figure}[h]
  \centering
  \subfigure[ExtraSensory (F1-socre $\uparrow$)]{\includegraphics[scale=0.26]{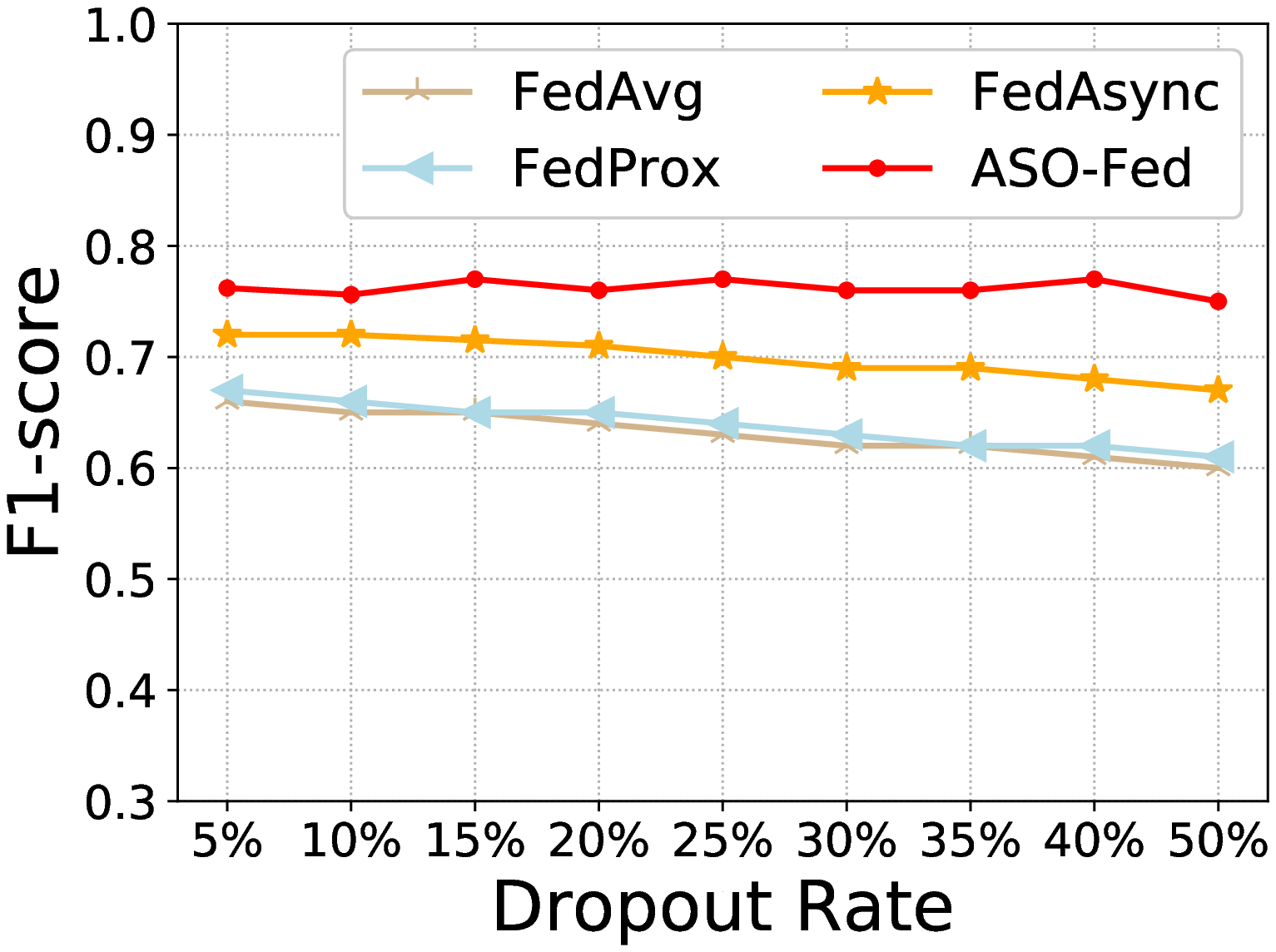}}
  \subfigure[Air Quality (SMAPE $\downarrow$)]{\includegraphics[scale=0.26]{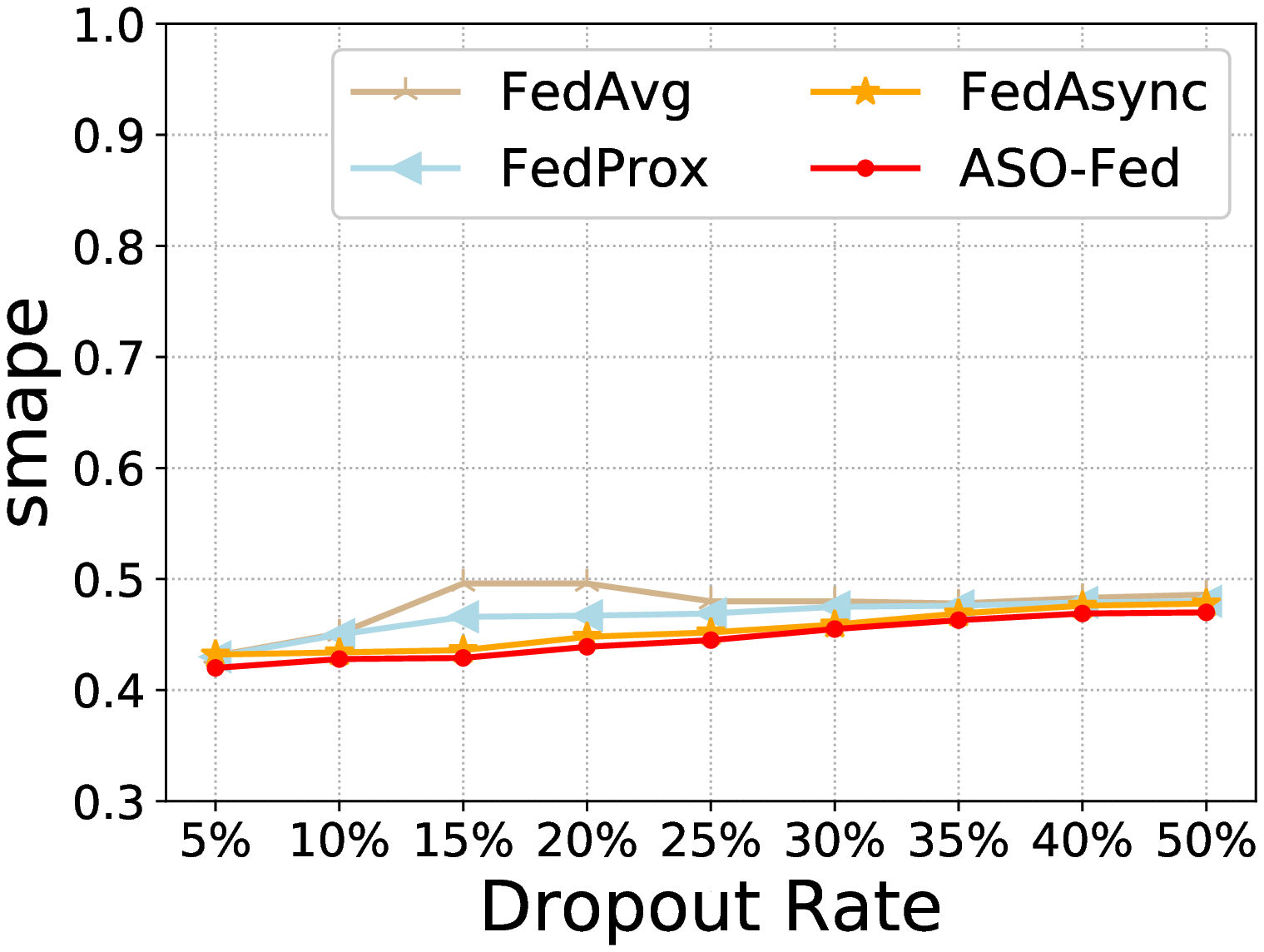}}
  \caption{Performance comparison of federated approaches as dropout rate of clients increases. \model~has better performance than the other federated frameworks. }\label{fig:dropout}
  \vspace{-1em}
\end{figure}

\begin{figure}[h]
  \centering
  \subfigure[ExtraSensory (F1-score $\uparrow$)]{\includegraphics[scale=0.26]{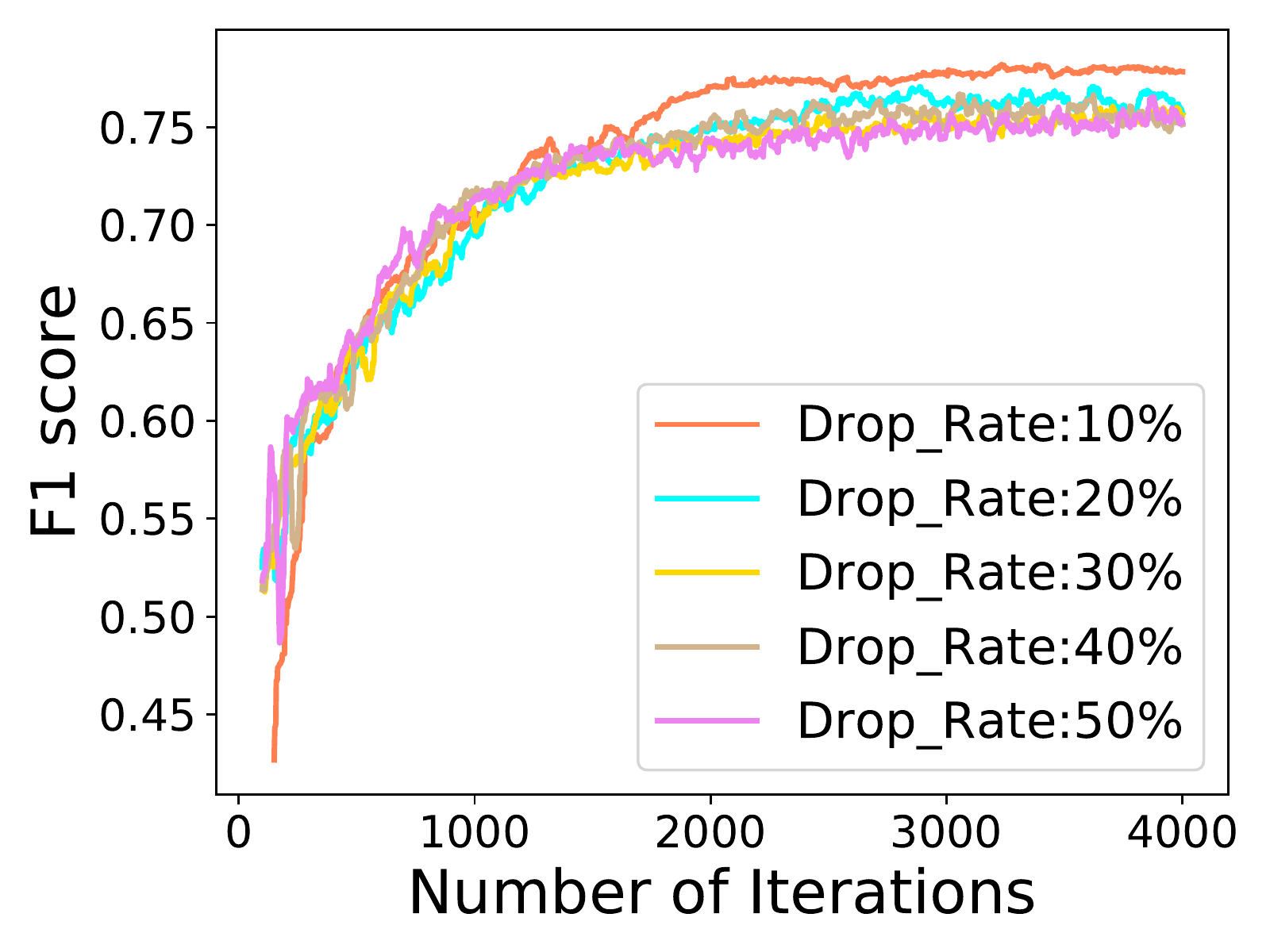}}
  \subfigure[Air Quality (SMAPE $\downarrow$)]{\includegraphics[scale=0.26]{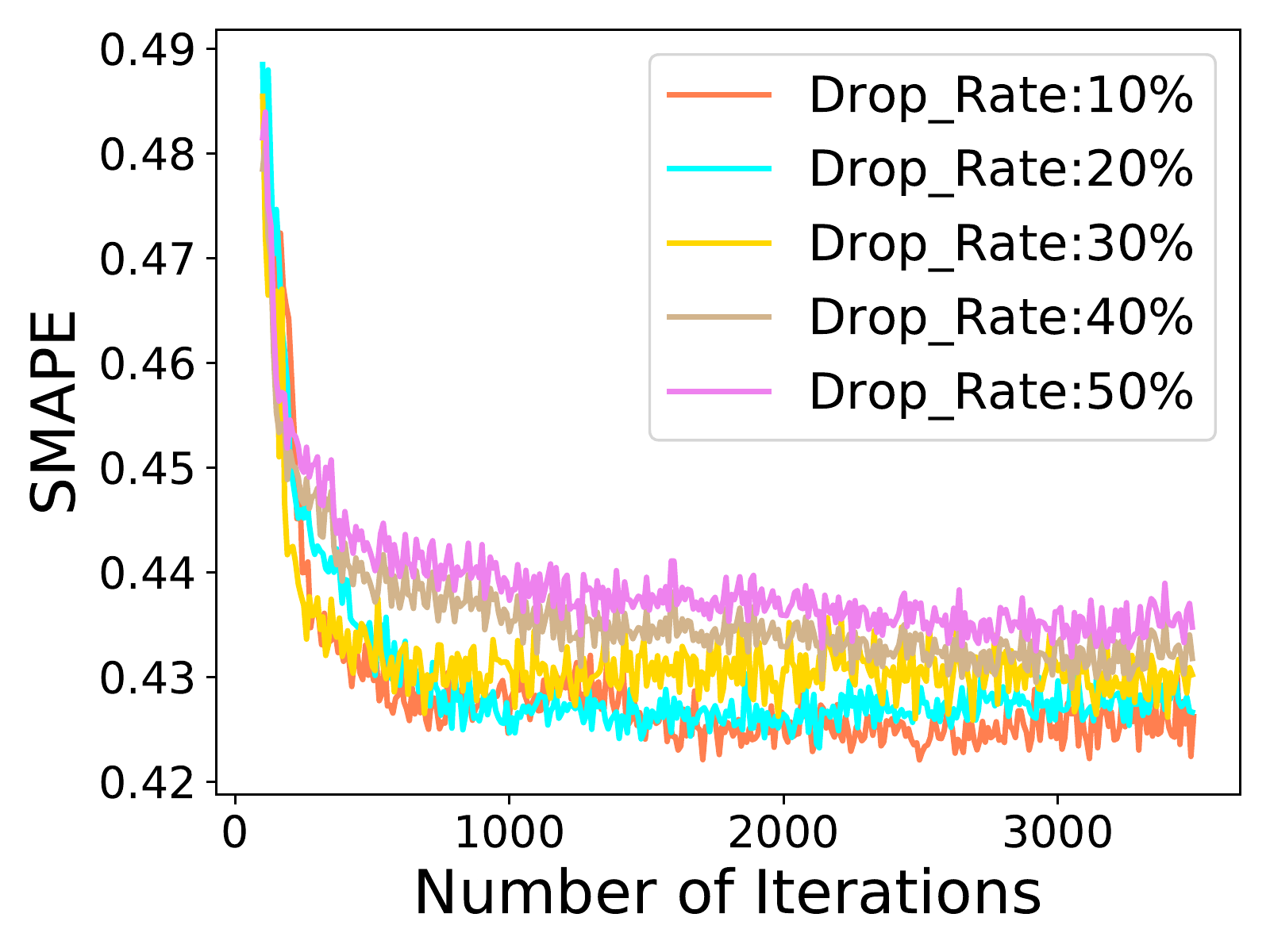}}
  \caption{The performance of \model~with clients periodically dropping out.  }\label{fig:dropout_curve}
\vspace{-1em}
\end{figure}

\noindent\textbf{Ablation Studies.} To evaluate the performance of central feature representation learning and local dynamic learning step size, we perform ablation studies with two additional models: \model(-F) and \model(-D).
From Table~\ref{table:performance} we notice that \model~outperforms \model(-F) by $1.06\% \sim 5.26\%$, which shows the effectiveness of central feature learning at generating a better feature representation across clients. We show the visualizations of learned features for the three real-world datasets in Section \ref{feature-visul}. \model(-D) has close prediction performance as \model, but as shown in Figure \ref{tab:prediction-curve}, \model(-D) needs a longer training time to converge than \model. This shows that local dynamic learning approach works effectively at lowering the overall computation cost.

\begin{figure*}[t]
 \centering
\includegraphics[trim={1.2cm 0.8cm 1cm 0.9cm},clip, width = 16cm]{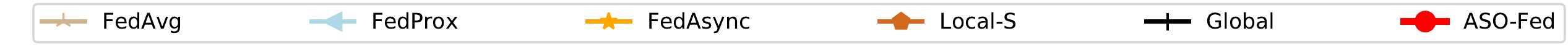}
\vspace{-1em}
\end{figure*}
\begin{figure*}[htbp]
  \centering
  \subfigure[FitRec (SMAPE $\downarrow$)]{\includegraphics[scale=0.27]{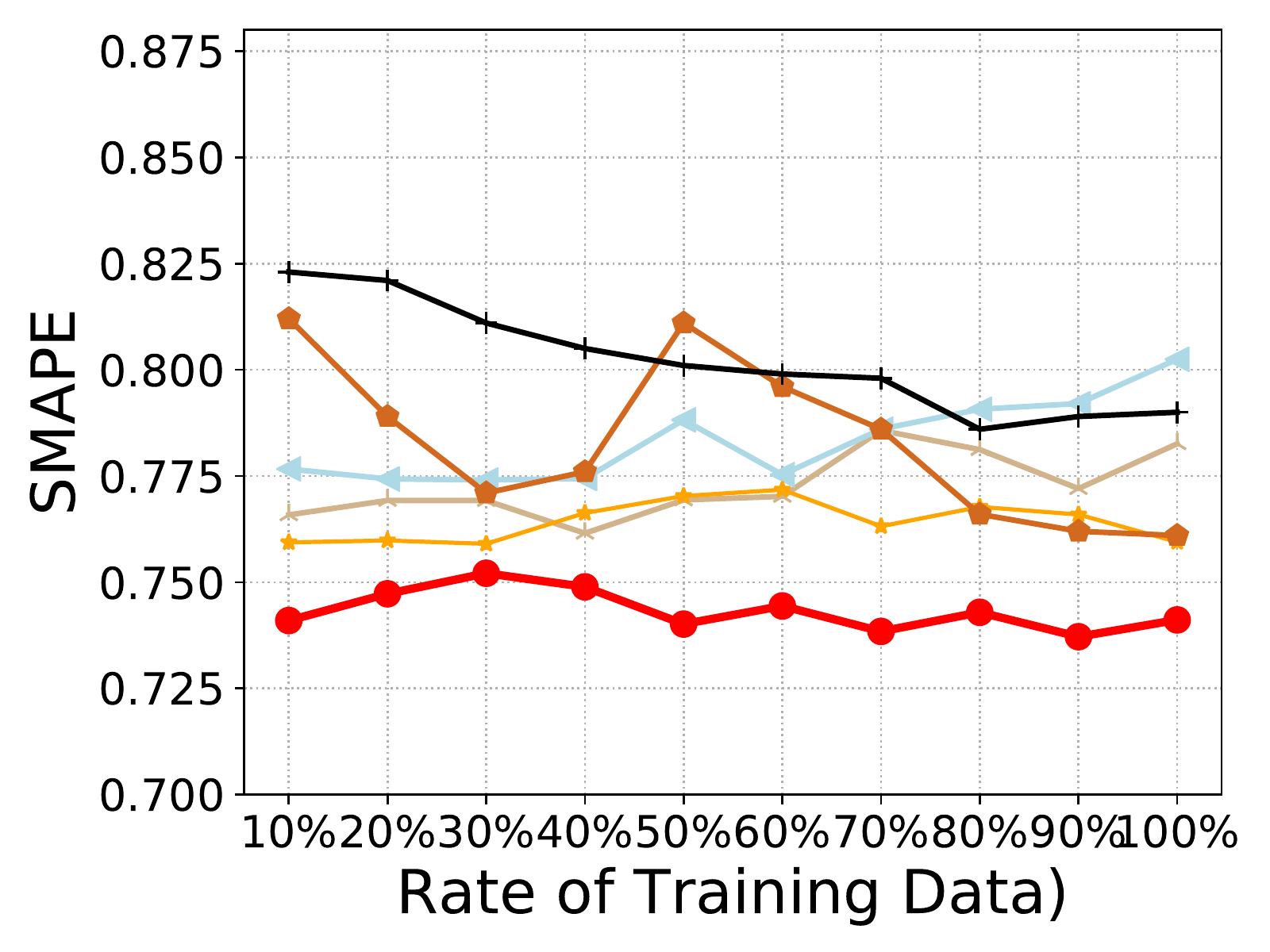}}
  \subfigure[Air Quality (SMAPE$\downarrow$)]{\includegraphics[scale=0.27]{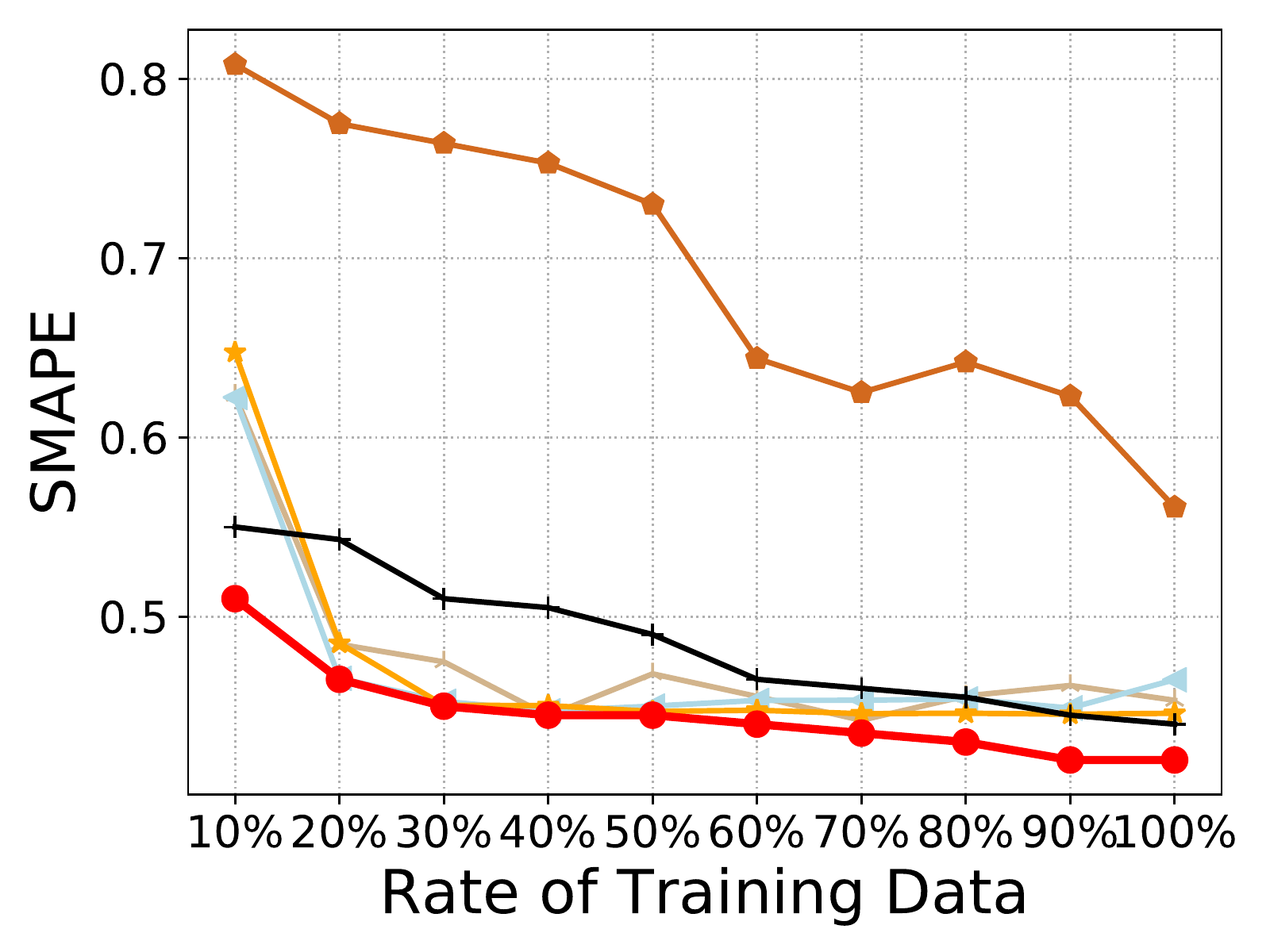}}
  \subfigure[ExtraSensory (F1-socre $\uparrow$)]{\includegraphics[scale=0.27]{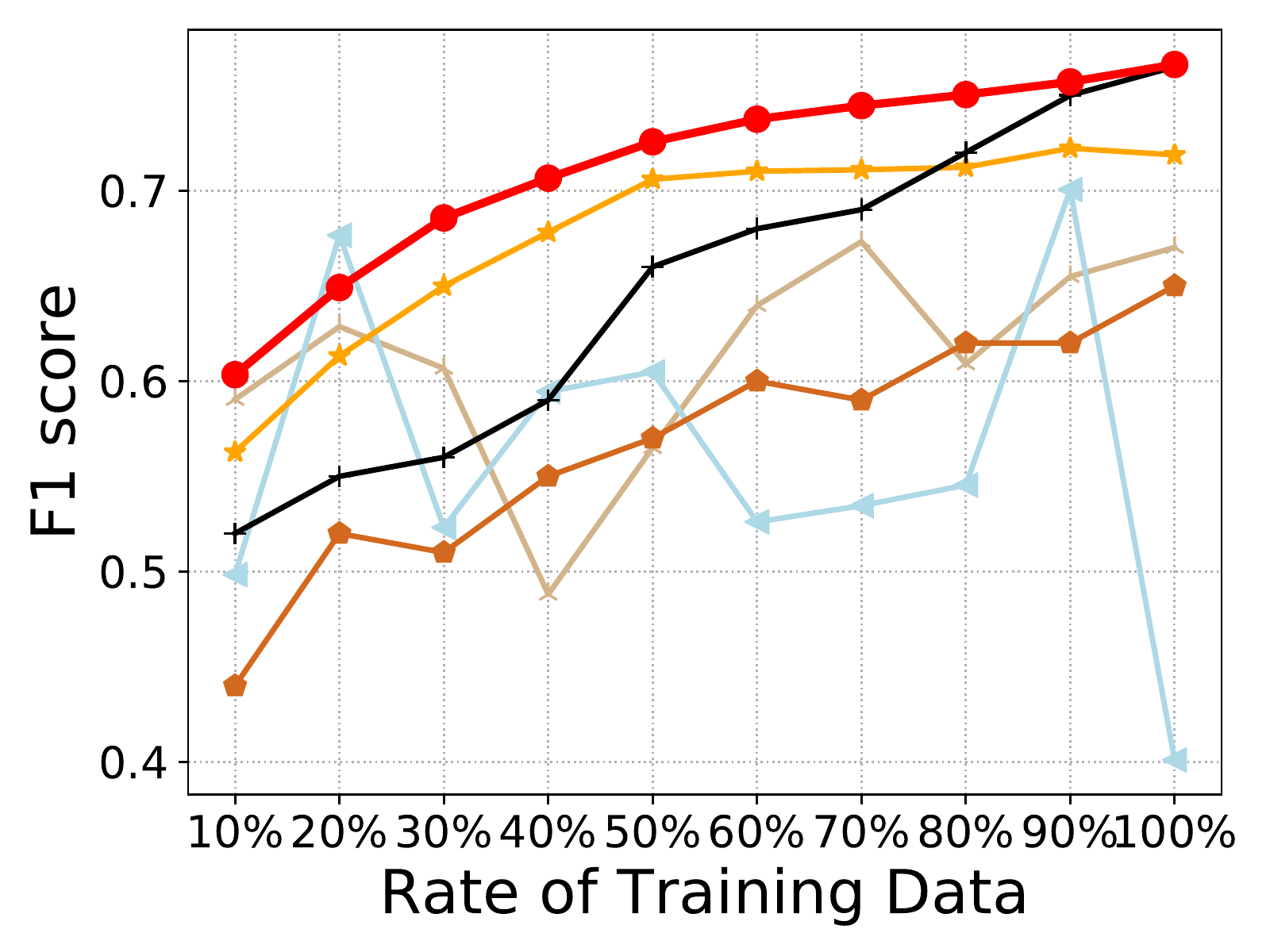}}
  \subfigure[Fashion-MNIST (Accuracy $\uparrow$)]{\includegraphics[scale=0.27]{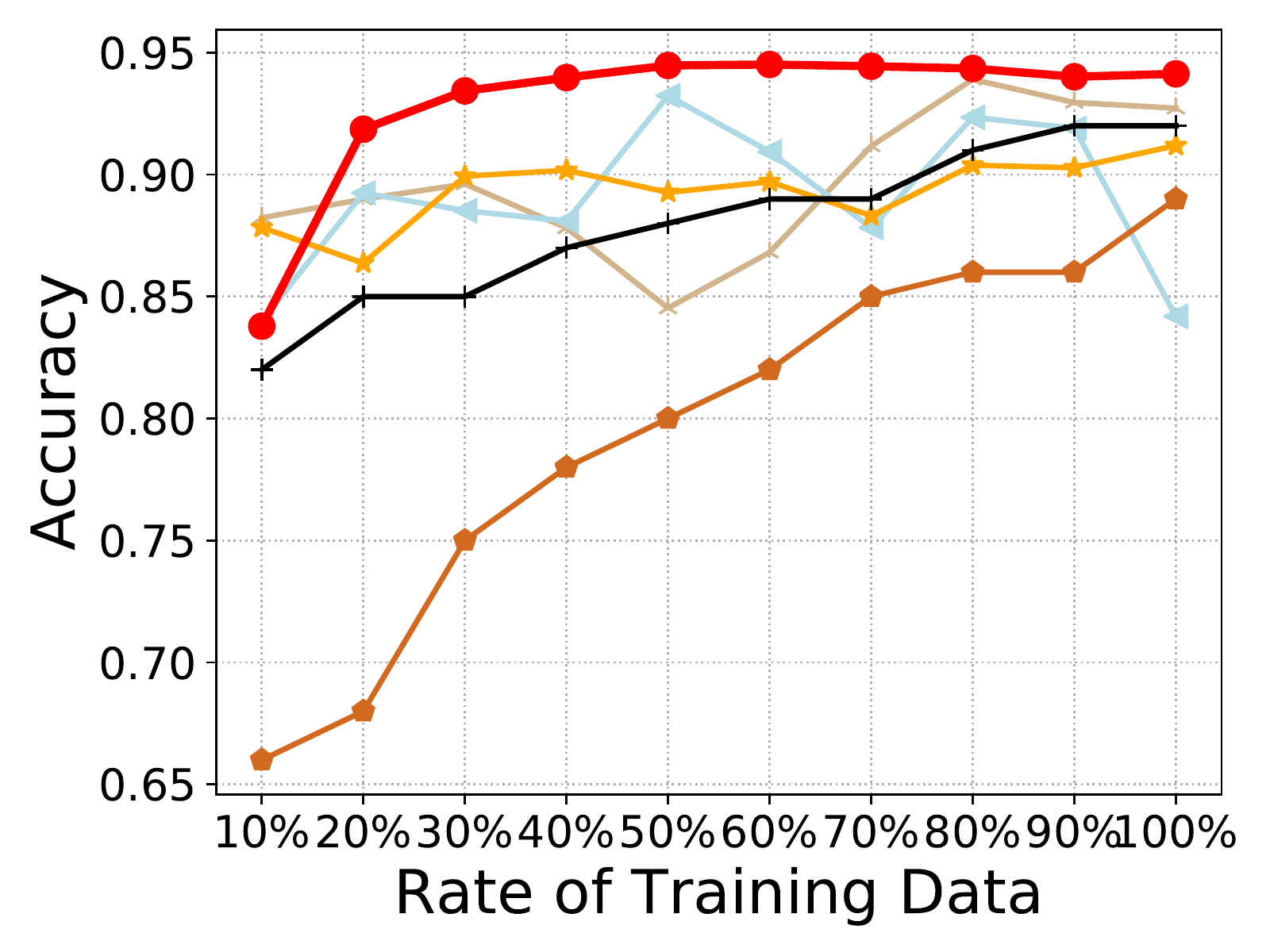}}
  \caption{Average performance comparison (SMAPE, F1, accuracy) on four datasets as training data increases. 
  }\label{tab:train-size}
  \vspace{-1em}
\end{figure*}

\subsection{Evaluation of Time Efficiency}

We report the running time of synchronous and asynchronous FL approaches to reach target test performance in Table~\ref{tab:timecost}.  
As seen from this table, FedAvg and FedProx have the highest computation cost across all four benchmarks. This is expected given that in synchronous protocol, the server aggregation has to wait for the slow client nodes to finish their computations. 
%
%
Among the comparison of asynchronous update models, \model~is more time efficient than \model(-D). FedAsync is close in running time as \model(-D). From the empirical results we note that the dynamic learning step size is a promising strategy and effective when there are significant delays in the network. \model~is close to \model(-F) in terms of computational cost except for the FitRec and ExtraSensory dataset. Because these two datasets have more complex features and additional computation for feature extraction requires more time, but \model~obtains better prediction performance with a little sacrifice of time efficiency.

\subsection{Robustness to Stragglers and Dropouts}

Stragglers are clients that lag in providing updates to the server due to a variety of reasons: communication bandwidth, computation load, and data variability.
%
%
%
%
In this Section, we investigate a common real-world scenario, when clients have no response during the entire training process or are not available in certain time windows. We refer to these non-contributing clients as dropouts.  

We set a certain portion of clients to be non-responsive. These clients will not participate in the training process. However, the reported results are evaluated on test data from all clients. 
Figure \ref{fig:dropout} shows the performance of federated learning approaches on Extrasensory classification and Air Quality regression benchmarks with an increasing fraction of clients being dropped from the learning process. 
As shown in Figure~\ref{fig:dropout}, for the ExtraSensory dataset, we observe that as the rate of dropout clients increases, the performance of the FedAvg model also drops. The same trend is noticed for FedProx. As for \model, the prediction performance is steady except for a slight decrease when the dropout rate exceeds $40\%$. Even when $50\%$ of clients are subject to dropout during training,  \model~can still achieve at least a 10\% improvement over synchronous FL models and FedAsync. For the Air Quality dataset, \model~has lower SMAPE errors than all other models as the dropout rate increases and the performance of \model~is relatively stable. However, as expected, if one of the nodes never sends updates to the central server, the model does not generalize. This explains the poor performance as the dropout rate increases.  

We also explore the performance effect of nodes periodically dropping and not providing updates to the server. 
To the best of our knowledge, no other methods for asynchronous federated learning with local streaming data directly address this issue. Therefore we do not draw comparisons to other methods. At each global iteration, we randomly select a certain fraction of clients as not participating in the training during the current iteration. Figure \ref{fig:dropout_curve} shows the convergence trend for different rates of periodically dropping clients. We notice that as the rate of periodically dropout clients increases, \model~still converges well with slightly worse performance. 
This shows that the performance of \model~is robust to relatively high rates of periodical dropouts.
%

\subsection{Results of Varying Training Samples}
To evaluate the incremental online learning process more explicitly, we display how the prediction performance changes with an increasing number of training samples in Figure~\ref{tab:train-size}. We perform experiments with different rates of all clients' training data and depict the average performance on all clients. From Figure \ref{tab:train-size}, for all datasets, \model~achieves the best performance with increasing rates of training data. Large fluctuations are observed in the results of FedAvg and FedProx, which exhibit an unstable model performance for synchronous approaches as the local data sets increase. Compared with the Global approach, asynchronous frameworks have more stable performance as the amount of training data increases, while the individual models learned by Local-S do not perform well. The analysis shows that that \model~learns an effective model with a smaller portion of training data. With increasing local data, \model~still outperforms the other competitors. 

\begin{figure}[h]
	\centering
	\includegraphics[trim={6cm 2.2cm 9.5cm 0.5cm},clip, width = 7.5cm]{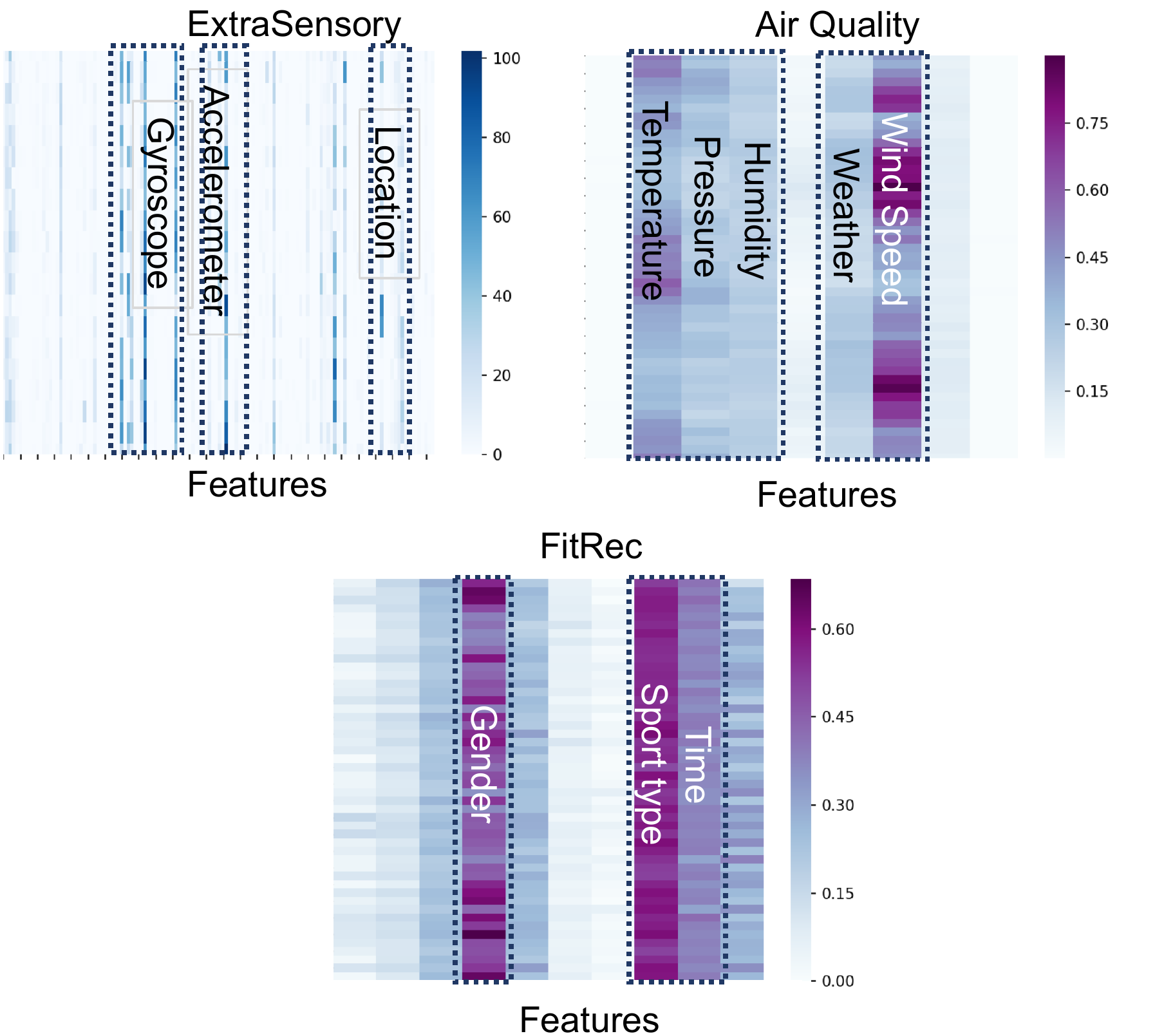}
	\caption{Feature representation learned on the server of three real-world datasets. Each column is the weights vector within 48 time steps over the input series.}\label{tab:feature}
\end{figure}

\subsection{Feature Representation learning}\label{feature-visul}

In this section, we present the qualitative results of the proposed feature representation learning on the server. In Figure~\ref{tab:feature}, we show the features learned from one client of three datasets respectively. For the client in ExtraSensory, the highlighted features are `Gyroscope', `Accelerometer' and `Location', and the corresponding labels are `walking', `at\_home'. 
For the client from Air Quality dataset, we observe that features with high weights are `Wind Speed' and `Temperature'. This makes sense given that the target values are air pollutants (e,g,. PM2.5, SO) and `Wind Speed' decides whether these pollutants can be dispersed. Air pollutants vary with seasons, and a higher concentration of air pollutants appears in winter time due to fuel consumption for heating in winter. Therefore `Temperature' is also a strong indicator for air pollutants. For the client from FitRec, the extracted features are `gender', `sport type' and `time'. Since the prediction targets are speed and heart rate, these three features have strong correlations with the targets. The above results show the effectiveness of feature learning in \model. 


\section{Conclusions and Future Work}
We propose a novel asynchronous online federated learning approach to tackle the learning problems on distributed edge devices. To the best of our knowledge, this is the first attempt to solve asynchronous federated learning with local streaming data.
Compared to synchronized FL approaches (FedAvg and FedProx), \model~is computationally efficient since the central server does not need to wait for lagging clients to perform aggregation. Compared with asynchronous approach (FedAsync), the proposed approach achieves better performance on all provided datasets, which indicates that the proposed asynchronous update method can better handle local streaming data. 
Time efficiency is compared on multiple benchmarks and the results show that the proposed \model~is faster than synchronized FL. 
The proposed \model~inherits the idea of using asynchronous update scheme as \cite{xie2019asynchronous,chen2019efficient}. Likewise, it shares the same communication bottleneck problem pointed out in \cite{kairouz2019advances}. Nevertheless, it is still an open issue to build communication efficient methods in asynchronous federated learning frameworks. Besides, feature learning component needs longer computation time when dealing with datasets which have complex features. Thus, there is a trade off between the prediction performance and computational costs. 
 In the future, we plan to develop communication efficient federated learning methods with asynchronous updating strategies.

\appendix

\section{Convergence Analysis}
\subsection{Proof of Lemma \ref{lemma_1}}

\begin{proof}
$F(w)$ is $\mu$\textit{-strongly convex}, we can get:
\begin{equation}
F(w') - F(w^t) \geq \langle \nabla F(w^t), w' - w^t \rangle + \frac{\mu}{2}||w'-w^t||^2,
\end{equation}
Let us define $C(w')$ such that:
\begin{equation}
C(w') = F(w^t) + \langle \nabla F(w^t), w' - w^t \rangle + \frac{\mu}{2} ||w'-w^t||^2,
\end{equation}
$C(w')$ is a quadratic function of $w'$, then it has minimal value when $\nabla C(w') = \nabla F(w^t) + \mu(w'-w^t) = 0$. Then the minimal value of $C(w')$ is obtained when $w' = w^t - \frac{\nabla F(w^t)}{\mu}$, which is:
\begin{equation}
C_{\text{min}} = F(w^t) - \frac{||\nabla F(w^t)||^2}{2\mu}, 
\end{equation}
For $F(w)$ is $\mu$-\textit{strongly convex}, we can complete the proof: 
\begin{equation}
F(w_*) \geq C(w_*) \geq C_{\text{min}} = F(w^t) -\frac{||\nabla F(w^t)|| ^2}{2\mu},
\vspace{-0.5em}
\end{equation}
\begin{equation}
2\mu(F(w^t) - F(w_*)) \leq  ||\nabla F(w^t)||^2.
\end{equation}
\end{proof}

\subsection{Proof of Theorem \ref{convex-model}}

\begin{proof}
With $F(w)$ is \textit{L-smooth}, we have: 
\begin{equation}
\begin{aligned}
&F(w^{t+1}) - F(w^t)
\leq \langle \nabla F(w^t), w^{t+1} - w^t \rangle + \frac{L}{2} ||w^{t+1}-w^t||^2 \\
&= -\nabla F(w^t)^\top \eta_k^t \frac{n'_k}{N'}\nabla\zeta_k(w^t) + \frac{L}{2}||\eta_k^t \frac{n'_k}{N'}\nabla\zeta_k(w^t)||^2,
\end{aligned}
\end{equation}
let $m = \eta_k^t \frac{n'_k}{N'} > 0$, with Assumption $1$ and local Bounded gradient dissimilarity, we can get:
\begin{equation}\label{Eq_1}
\begin{aligned}
&\mathbb{E}(F(w^{t+1})) - F(w^t)\\ 
&\leq -m\nabla F(w^t)^\top \mathbb{E}(\nabla\zeta_k(w^t)) + \frac{L m^2}{2}\mathbb{E}(|| \nabla\zeta_k(w^t)||^2)\\
&\leq -m\epsilon || \nabla F(w^t) ||^2 + \frac{L m^2V^2}{2} ||\nabla F(w^t)||^2 \\
& = -m(\epsilon - \frac{L mV^2}{2})||\nabla F(w^t)||^2,
\end{aligned}
\end{equation}
We can easily prove that $-m(\epsilon-\frac{L m V^2}{2})$ is monotonically increasing while $m>0$. Since $n'_k < N'$, thus $m = \eta_k^t \frac{n'_k}{N'} < \eta_k^t$. Then we have $-m(\epsilon-\frac{L m V^2}{2}) < -\eta_k^t(\epsilon-\frac{L \eta_k^t V^2}{2})$. 

With Lemma \ref{lemma_1}, and let $\gamma = \epsilon-\frac{L \eta_k^t V^2}{2} $, we can rewrite Equation (\ref{Eq_1}) as:
\begin{equation}\label{Eq_2}
\begin{aligned}
\mathbb{E}(F(w^{t+1})) - F(w^t) &\leq -2\mu \eta_k^t\gamma (F(w^t) - F(w_*)), 
\end{aligned}
\end{equation}
Then we move $F(w^t)$ on left side to right and subtract $F(w_*)$ from both sides, and get:
\begin{equation}\label{Eq_3}
\begin{aligned}
\mathbb{E}(F(w^{t+1})) - F(w_*) &\leq (1-2\mu \eta_k^t\gamma) (F(w^t) - F(w_*)),
\end{aligned}
\end{equation}
Since $\bar\eta_k < \eta_k^t$, then by taking expectation of both sides, and telescoping, we have:
\begin{equation}\label{Eq_proof}
\begin{aligned}
\mathbb{E}(F(w^{t+1}) - F(w_*)) &\leq (1-2\mu\bar\eta_k \gamma')(\mathbb{E} (F(w^t)- F(w_*)).
\end{aligned}
\end{equation}
When $t+1 = T$, the above inequality becomes Equation~(\ref{Eq_4}). Thus we complete the proof.
\end{proof}

\subsection{Proof of Theorem \ref{nonconvex-model}}

\begin{proof}
With $m<\eta_k^t$ and $\gamma = \epsilon-\frac{L \eta_k^t V^2}{2}$, we replace $m$ with $\eta_k^t$ and take the full expectation of Equation (\ref{Eq_1}), we have:
\begin{equation}\label{Eq_5}
\begin{aligned}
\mathbb{E}(F(w^{t+1})) &\leq \mathbb{E}(F(w^t))-\eta_k^t\gamma \mathbb{E}(||\nabla F(w^t)||^2),
\end{aligned}
\end{equation}
Then summing up (\ref{Eq_5}) over global iteration $T$, we can get:
\begin{equation}\label{Eq_6}
\begin{aligned}
\mathbb{E}(F(w^{t+1})) &\leq F(w^0)- \sum_{t=0}^{T-1} \eta_k^t\gamma \mathbb{E}(||\nabla F(w^t)||^2),
\end{aligned}
\end{equation}
From Assumption \ref{assumption1}.1 we can get $F(w_*) \leq \mathbb{E}(F(w^{t+1}))$, then we have:
\begin{equation}\label{Eq_7}
\begin{aligned}
F(w_*) &\leq F(w^0)- \sum_{t=0}^{T-1} \eta_k^t\gamma \mathbb{E}(||\nabla F(w^t)||^2),
\end{aligned}
\end{equation}
If we set $\eta_k^t < \frac{2\epsilon-1}{L V^2} \leq max({r_k^t}\bar\eta) = \bar\eta$, we can get $\eta_k^t(\epsilon-\frac{L \eta_k^t V^2}{2}) > \frac{\eta_k^t}{2}$. Rearrange Equation (\ref{Eq_7}) we can get:
\begin{equation}\label{Eq_8}
\begin{aligned}
\sum_{t=0}^{T-1} \frac{\eta_k^t}{2}\mathbb{E}(||\nabla F(w^t)||^2) &\leq \sum_{t=0}^{T-1} \eta_k^t(\epsilon-\frac{L \eta_k^t V^2}{2}) \mathbb{E}(||\nabla F(w^t)||^2)\\
&\leq F(w^0)- F(w_*).
\end{aligned}
\end{equation}
Then we get Equation (\ref{Eq_nonconvex}) and complete the proof.
\end{proof}




\bibliographystyle{IEEEtran}
\bibliography{bibfile}






\end{document}